\documentclass[fleqn,usenatbib]{mnras}

\usepackage{newtxtext,newtxmath}

\usepackage[T1]{fontenc}
\usepackage{ae,aecompl}

\usepackage{verbatim}
\usepackage{graphicx}	
\usepackage{amsmath}	
\usepackage{enumitem}
\usepackage{array}
\usepackage{multirow}
\usepackage{lipsum}
\usepackage{siunitx} 
\DeclareSIUnit\solarmass{M\ensuremath{_\odot}}
\usepackage{subfiles}
\usepackage{diagbox}
\usepackage{gensymb}

\newcommand{\fnsm}{\ensuremath{f_{\rm NSM}}}

\newcommand{\Rnsm}{\ensuremath{R_{\rm NSM}}}
\newcommand{\prompt}{\textit{prompt}}

\newcommand{\ansm}{\ensuremath{\alpha_{\rm NSM}}}
\newcommand{\kalpha}{\ensuremath{k_\alpha}}
\newcommand{\starf}{\textit{star-forming}}
\newcommand{\passive}{\textit{passive}}

\newcommand{\sunmass}{\SI{}{\solarmass}}

\title[On the hosts of neutron star mergers in the nearby Universe]{On the hosts of neutron star mergers in the nearby Universe}

\author[L. Cavallo et al.]{
L. Cavallo,$^{1}$ L.Greggio$^2$
\\
$^{1}$Physics and Astronomy Department Galileo Galilei, University of Padova, Vicolo dell’Osservatorio 3, I–35122, Padova, Italy\\
$^{2}$INAF - Osservatorio Astronomico di Padova, Vicolo dell'Osservatorio 5, Padova, 35122 Italy
}

\date{Accepted XXX. Received YYY; in original form ZZZ}

\pubyear{2020}

\begin{document}
\label{firstpage}
\pagerange{\pageref{firstpage}--\pageref{lastpage}}
\maketitle

\begin{abstract}
Recently, the characterisation of binary systems of neutron stars has become central in various fields such as gravitational waves, gamma-ray bursts (GRBs), and the chemical evolution of galaxies. In this work, we explore possible observational proxies that can be used to infer some characteristics of the delay time distribution (DTD) of neutron star mergers (NSMs). We construct a sample of model galaxies that fulfils the observed galaxy stellar mass function, star formation rate versus mass relation, and the cosmic star formation rate density. The star formation history of galaxies is described with a log-normal function characterised by two parameters: the position of the maximum and the width of the distribution.
We assume a theoretical DTD that mainly depends on the lower limit and the slope of the distribution of the separations of the binary neutron stars systems at birth. We find that the current rate of NSMs ($\mathcal{R}=320^{+490}_{-240}$ Gpc$^{-3}$yr$^{-1}$) requires that $\sim0.3$ per cent of neutron star progenitors lives in binary systems with the right characteristics to lead to a NSM within a Hubble time. We explore the expected relations between the rate of NSMs and the properties of the host galaxy. We find that the most effective proxy for the shape of the DTD of NSMs is the current star formation activity of the typical host. At present, the fraction of short-GRBs observed in star-forming galaxies favours DTDs with at least $\sim40\%$ of mergers within $100$ Myr. This conclusion will be put on a stronger basis with larger samples of short-GRBs with host association (e.g. $600$ events at $z \leq 1$). 
\end{abstract}



\defcitealias{G13}{G13} 
\defcitealias{Peng2010ApJ...721..193P}{P10} 

\section{Introduction} \label{sec:intro}
In the last decade, the coalescence of binary systems of neutron stars has gained a lot of interest in multiple fields such as multi-messenger astrophysics, nucleosynthesis studies, chemical evolution of galaxies, and high-energy astrophysics.
The first observation of a Neutron Star Merger (NSM) has been obtained with the gravitational wave event GW170817 \citep{Abbott2017}, which has been localised in a sky region of about 30 deg$^2$. This observation has confirmed that these events are associated with a gravitation wave signal. After $\sim11$ hours, several research groups \citep{Abbott2017,Coulter2017,Soares-Santos2017ApJ...848L..16S, Valenti2017ApJ...848L..24V} have independently detected the optical counterpart of GW170817, the kilonova AT2017gfo. In the following days, the light from the kilonova and afterglow emission have been observed and analyzed \citep[see][]{Kilo2017ApJ...848L..12A, Villar2017ApJ...851L..21V, Hajela2019ApJ...886L..17H, Troja2019MNRAS.489.1919T}. For many years ejecta of neutron star mergers (NSMs) have been indicated as a strong source of rapid neutron-capture (r-)process
elements \citep[][]{Symbalisty1982ApL....22..143S, Freiburghaus1999ApJ...525L.121F, Rosswog2000A&A...360..171R, Oechslin2007A&A...467..395O, Panov2008AstL...34..189P, Perego2014MNRAS.443.3134P,Rosswog2014MNRAS.439..744R, Wanajo2014,Eichler2015, Goriely2015MNRAS.452.3894G, Lippuner2017MNRAS.472..904L, Thielemann2017}. In this regard, the evolution of the light curve of the kilonova of the GW170817 event suggests a significant production of r-process elements \citep[][]{Cowperthwaite2017ApJ...848L..17C,Tanaka2017PASJ...69..102T, Villar2017ApJ...851L..21V}. This reinforces the theory that rapid neutron capture processes occur in this type of system.
In the following days, the gravitational waves (GW) emission from GW170817 has been correlated with a space and time coincident short gamma-ray burst (SGRB), GRB 170817A \citep{AbbottGRB2017ApJ...848L..13A, Goldstein2017ApJ...848L..14G, Savchenko2017ApJ...848L..15S}. The detection of GRB170817A $\sim2$ seconds after the GW event has reinforced the hypothesis that NSM (and in general compact binary mergers) are the progenitors of SGRBs \citep{Eichler1989Natur.340..126E,Giacomazzo2013ApJ...762L..18G,Tanvir2013Natur.500..547T,Berger2014ARA&A..52...43B}. In the following years, a second NSM event (GW190425) \citep[][]{Abbott2020} has been detected, as well as two neutron star-black hole mergers: GW190426 and GW190814 \citep{Abbott2021ApJ...915L...5A}. Unfortunately, for these three events, the observation of the electromagnetic counterparts has not been reported.
In the last two years, observations have continued with the third observing run (O3) but no new NSM event has been detected; as a consequence  \citet{2021ApJ...913L...7A} have updated the present-day cosmic rate of NSM to $\mathcal{R}=320^{+490}_{-240}$ Gpc$^{-3}$, yr$^{-1}$, that is about $\sim 1/3$ of the one previously derived by \citet{Abbott2020}, i.e. $\mathcal{R}=1090^{+1720}_{-800}$ Gpc$^{-3}$ yr$^{-1}$. \\
As confirmed by the kilonova AT2017gfo, NSMs are producers of rapid neutron-capture elements. Their impact on the evolution of r-process elements has been investigated with chemical evolution models. In these studies, the production time scale of a certain element plays a key role. For r-process elements produced by NSMs, the production time-scale is linked to the delay time, i.e. the time between the birth of the primordial binary system and its final merging. For this reason, the chemical evolution of galaxies can provide information which put some constraints on the delay times of NSMs. An element such as europium (Eu) is often used as a good tracer of the r-process, mostly because it appears that $\sim 90\%$ of the solar Eu has been produced via rapid neutron capture processes \citep[][]{Howard1986ApJ...309..633H,Bisterzo2015MNRAS.449..506B}. This conclusion follows from the insufficient contribution to Europium of the r-process in intermediate mass stars, as estimated on the basis of current stellar models. In the last years, a series of works have shown that the Eu enrichment should take place on short timescales \citep[][]{Matteucci2014MNRAS.438.2177M, Cescutti2015A&A...577A.139C, Ishimaru2015ApJ...804L..35I,Cote2019ApJ...875..106C, cavallo2021MNRAS.503....1C}. In particular,
it has been argued that NSMs can be the sole source of Eu in the Galaxy only if the gravitational delay of NSMs  (i.e. the time between the formation of the binary neutron star system and their merging event) is short (up to $\sim 100$ Myr). On the other hand, if NSMs have relatively long gravitational delays, in order to explain both the spread and the average trend of [Eu/Fe] observed in our Galaxy, some additional site of Eu production with short timescales must be invoked, e.g. Core Collapse (CC) SNe. \\
\begin{figure*}
    \centering
    \includegraphics[trim={0cm 0cm 0cm 0cm}, clip, width=0.9\textwidth]{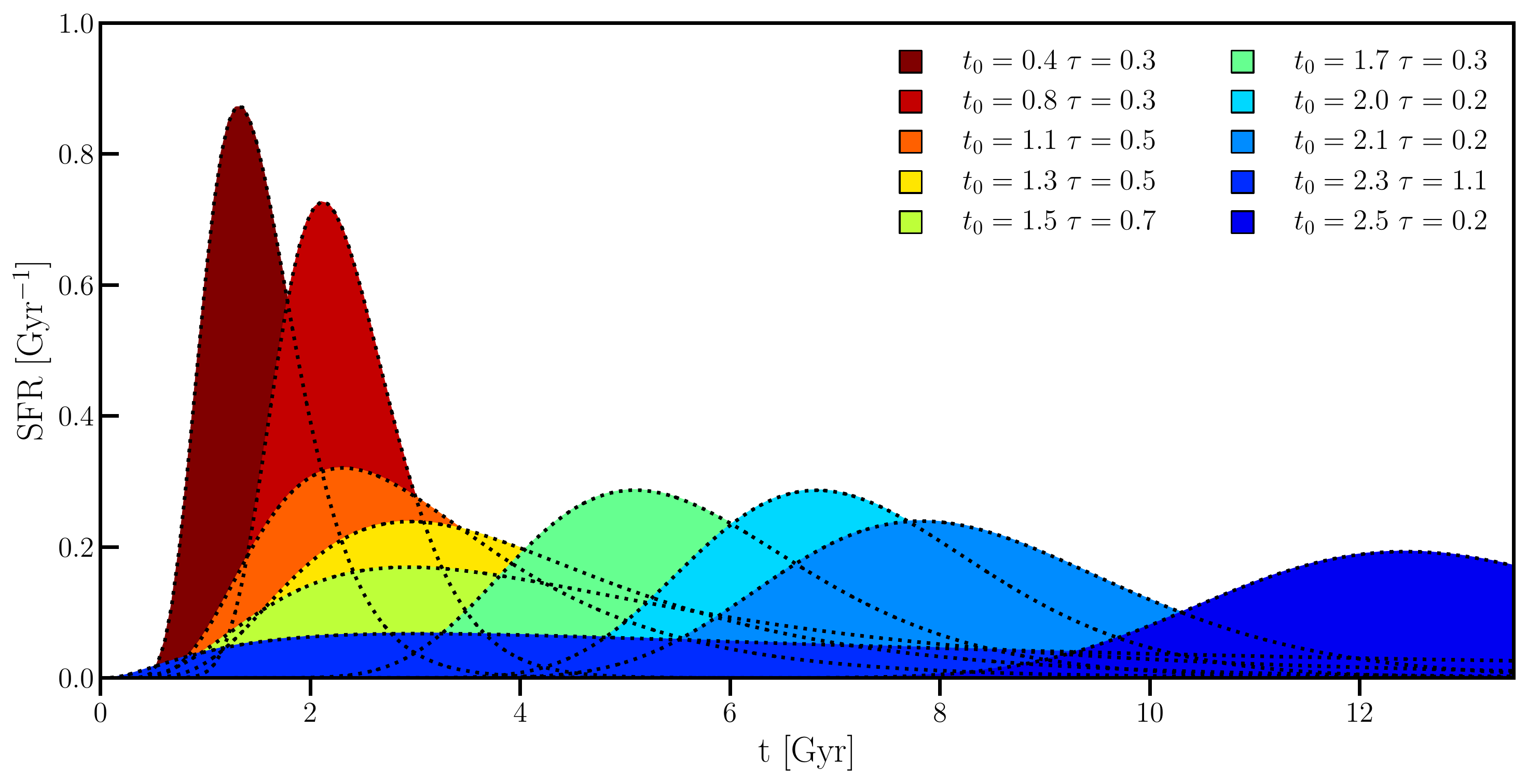}
    \caption{Selection of SFH laws for galaxies in the sample of \citetalias{G13}. Different colors correspond to different combinations of the ($t_0, \tau$) parameters as labelled.} 
    \label{fig:SFH}
\end{figure*}
\begin{figure}
    \centering
    \includegraphics[trim={0cm 0cm 0cm 0cm}, clip,width=0.5\textwidth]{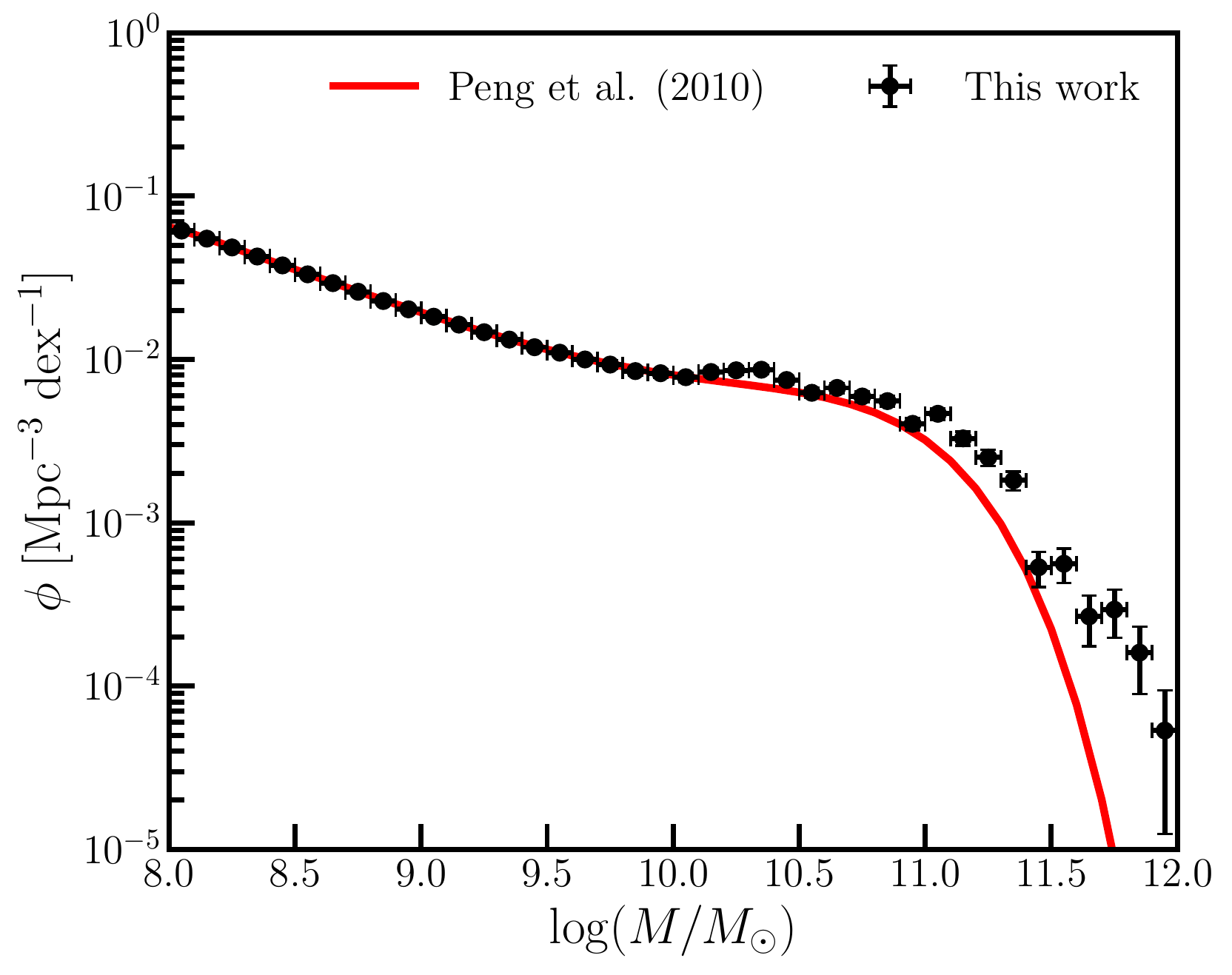}
    \caption{Galaxy stellar mass functions of \citetalias{Peng2010ApJ...721..193P} (red line) and the one of our sample (black dots). The horizontal bars show the galaxy mass bins, the vertical bars the Poissonian uncertainty on the number of galaxies in the bin. Note that we refer to the \citet{Peng2010ApJ...721..193P} GSMF rescaled to the Salpeter IMF.}
    \label{fig:MDF}
\end{figure}
The optical counterpart of GW170817 has allowed the identification of its host galaxy: NGC 4993. This galaxy is located at a distance of $39.5$ Mpc, in agreement with the distance of GW170817 ($24-48$ Mpc) estimated by \citet{Abbott2017} from the analysis of the GW signal. NGC4993 is an early-type galaxy that presents a low current star formation rate. Thus, in principle, the progenitor of GW170817 could be a young system. However, by analysing the local environment around the kilonova AT2017gfo \citet{Kilpatrick_2022} have not found clear evidence of recent star formation. This indicates that the progenitors of GW170817 more likely arise from an old stellar population, implying that at least some NSMs should have long delay times.\\
As highlighted above, the observation of GW170817 in a galaxy with a low current SFR supports long delay times for NSMs; on the other hand, chemical evolution models require NSMs with short delay times. The assumption of a delay time distribution (DTD) for NSMs satisfies both requests. This hypothesis is supported by the observations of SGRBs. As already mentioned, SGRBs are thought to be related to NSMs; thus, the census of SGRBs could provide an alternative way to constrain the properties of the DTD of NSMs. The analysis of observed SGRBs suggests that the DTD of SGRBs scales with the inverse of the delay time \citep[see][]{Guetta2006A&A...453..823G,Davanzo2014MNRAS.442.2342D, Ghirlanda2016}. \\
In a simple approach, DTDs are often assumed as pure power-laws with a certain slope (i.e. $f_{\rm NSM}(t)\propto t^{-s}$); typically $0.5\leq s\leq2$. However, as pointed out by \citet{Simonetti2019MNRAS.486.2896S} and \citet{Greggio2021}, similar to the case of SNe Ia, the DTD should be characterised by an early wide peak, with a width equal to the difference between the evolutionary lifetimes of the least and most massive neutron star progenitors. Furthermore, the slope of the DTD at late epochs should be a function of the shape of the distribution of separations of the NS-NS systems at birth. The DTD of NSMs can also be computed numerically with binary population synthesis (BPS) models \citep[][]{GiacobboMapelli2018MNRAS.480.2011G, Belczynski2020A&A...636A.104B, Tang2020MNRAS.493L...6T}, which follow the evolution of individual systems up to the merging event. To do that, these models include many ingredients some of which are not well constrained, e.g. the initial-final mass relations, the efficiency of the common envelope (CE) phase, the initial distribution of mass ratios and separations, the SN kick and its effects on the binary system, and the possible dependence of these properties on the chemical composition. The resulting DTD is thus subject to some weaknesses while being at the same time a rigid prediction. \\
As an alternative, \citet{Greggio2021} proposed a DTD for NSMs expressed by a simple analytical formulation identifying some physical parameters which mostly control the shape of the DTD.
In particular, following the same method used in \citet{Greggio2005A&A...441.1055G} for the DTD of SNIa's, they developed a parametrized DTD which depends on the distribution of three key parameters of the neutron star binaries at birth: their separations, the total mass of the system, and orbit eccentricity.\\
In this paper, we investigate possible observational facts that can be used to constrain the main characteristics of the DTD of NSMs. To do that, we build a sample of galaxies that complies with the major observational trends: i) the galaxy stellar mass function (GSMF) observed for nearby galaxies \citep[see][hereafter \citetalias{Peng2010ApJ...721..193P}]{Peng2010ApJ...721..193P}; ii) the star formation rate density (SFRD) obtained by \citet{MD2014}; iii) the star-forming main sequence of galaxies as derived by \citet{Renzini2015ApJ...801L..29R}. We assume that the SFH of these galaxies can be expressed as a log-normal function that depends on the epoch of the peak ($t_0$) and on the width of the function ($\tau$) \citep[see][hereafter \citetalias{G13}]{G13}. To calculate the rate of NSMs in the Universe, besides the SFH of the galaxies, we need the DTD, for which we adopt the models developed by \citet{Greggio2021}.
These ingredients allow us to compute models for the trend of NSM rate as a function of redshift; in order to derive the actual value of the rate we need to determine a scaling factor (\ansm) which we calibrate by imposing that the models 
reproduce the present rate of NSMs determined by \citet{2021ApJ...913L...7A} ($\mathcal{R}=320^{+490}_{-240}$ Gpc$^{-3}$ yr$^{-1}$). \\
Based on our models, we then compute the fraction of NSMs hosted by \starf\ galaxies classified on the basis of their specific SFR (sSFR) according to different criteria, and compare it with the fraction of SGRBs observed in \starf\ galaxies derived by \citet{Fong2022barXiv220601764N}.\\
The paper is organised as follows. In Section \ref{sec:Kilonovaerates} we describe the sample of galaxies and the DTDs used in the computation. In Section \ref{sec:zresults} we present the redshift evolution of the NSMs rate and the calibration of \ansm\ . In Section \ref{sec:fractions} we present the demographic of observed SGRBs, and compare the data to some predictions of our model. We discuss our results in Section \ref{sec:Discussion} and draw some conclusion in Section \ref{sec:conclusions}. In this work we assume a $\Lambda$CDM model with ${\rm H}_0 = 70$ km s$^{-1}$ Mpc$^{-1}$, $\Omega_{\rm M}=0.3$, and $\Omega_{\Lambda}=0.7$.
\begin{figure*}
    \centering
    \includegraphics[trim={0cm 0cm 0cm 0cm}, clip, width=0.8\textwidth]{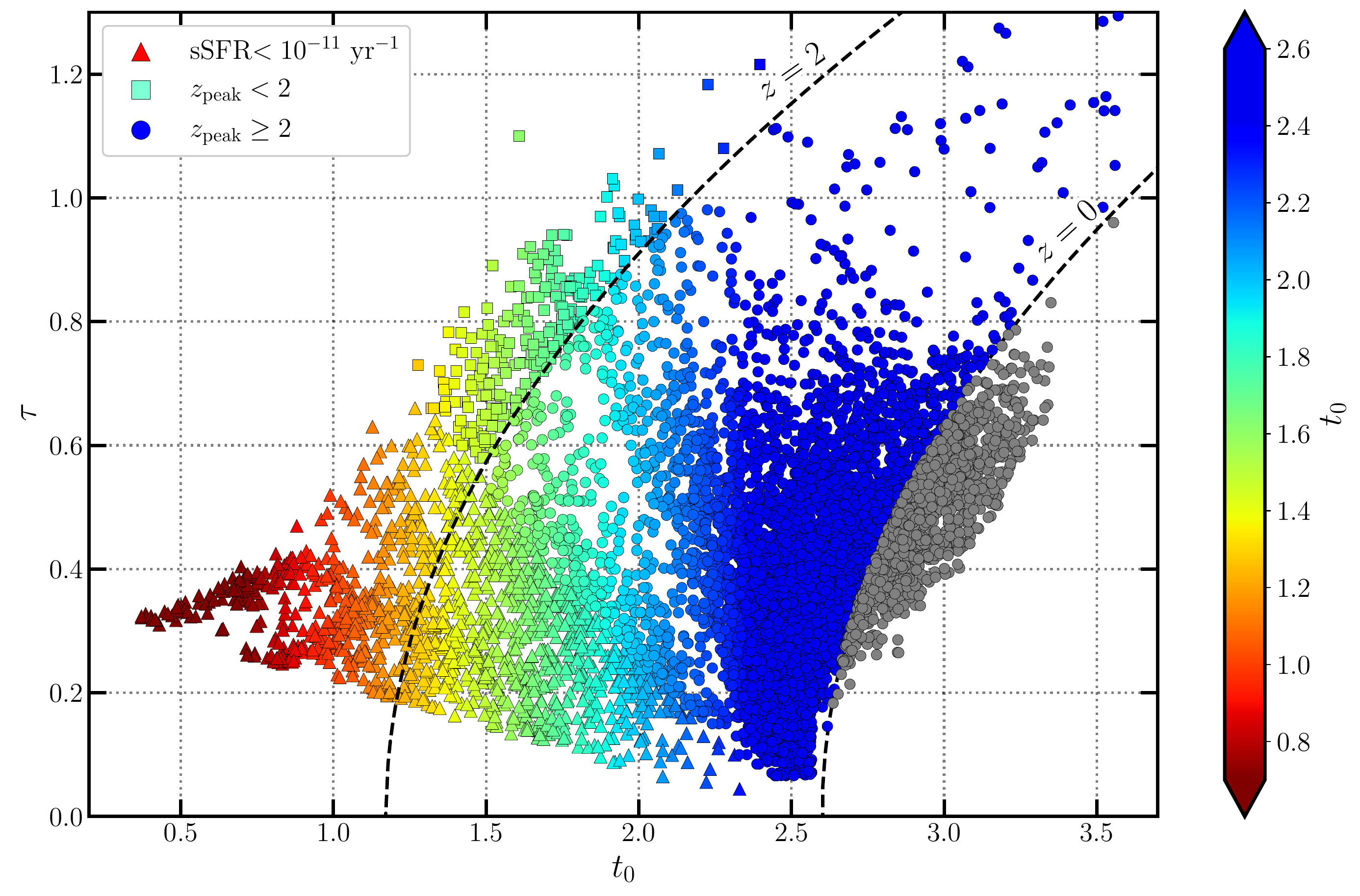}
    \caption{Distribution of the parameters $t_0$ and $\tau$ characterizing the log-normal SFH laws. Symbols are individual galaxies colour-coded as reported by the colour map on the right. Different colours indicate the value of $t_0$. Different symbols are used to distinguish between: i) galaxies with a specific star formation rate (sSFR) at $z=0$ below $10^{-11}$ yr$^{-1}$ (triangles); ii) galaxies with high sSFR (greater than $10^{-11}$ yr$^{-1}$) and with an early SFR-peak ($z_{\rm peak}>2$, squares); iii) galaxies with high sSFR and a late SFR-peak ($2\leq z_{\rm peak} \leq 0$, circles); iv) galaxies whose SFR-peak has not yet occurred (grey circles). Black dashed lines show the relation between the $\tau$ and $t_0$ parameters for SFHs that peak at $z_{\rm peak} = 2$ and 0.}
    \label{fig:colorCode}
\end{figure*}
\section{Neutron star merger rate} \label{sec:Kilonovaerates}
The DTD function is the distribution of the delay times for a stellar population of unitary mass, formed by an instantaneous burst of star formation. The DTD is a crucial ingredient for the computation of NSM rates and the chemical evolution of elements released to the interstellar medium (ISM) by these events. Indeed, the NSM rate in a galaxy at a given time $t$ can be computed as:
\begin{equation}\label{eqn:rate}
    \Rnsm(t) = \kalpha \, \ansm \,\int_{0}^{t} \psi(t-\tau) \fnsm(\tau)  d\tau
\end{equation}
where \kalpha\ is the number of NS progenitors per unit mass in a single stellar generation, \ansm\ is the fraction of them with the right characteristics to lead to a NSM, $\psi(t)$ is the star formation history (SFH, in $\sunmass {\rm yr}^{-1}$) of the galaxy, and $\fnsm(\tau)$ is the DTD function (see, e.g., \citet{Simonetti2019MNRAS.486.2896S}). The parameter \kalpha\ depends on the assumed initial mass function (IMF) and on the mass range of NS progenitors. It can be calculated as:
\begin{equation}
    \kalpha = \int_{m_1}^{m_2} \phi(m)dm
\end{equation}
where $m_1$ and $m_2$ are the minimum and maximum mass of NS progenitors, respectively, and $\phi(m)$ is the assumed IMF normalized to a unitary mass over the total stellar mass range. In this work we adopt: a \citet{Salpeter1955ApJ...121..161S} IMF, a total range for stellar masses from 0.1 to 120 \sunmass, and NS progenitors ranging from 9 to 50 $\sunmass$ . With these assumptions we get \kalpha$\simeq\num{6e-3}$ $\sunmass^{-1}$. The upper limit to the mass of NS progenitors is chosen according to a constraint from chemical evolution models \citep[]{Matteucci2014MNRAS.438.2177M}. With this formalism the parameter \ansm\ acts like a normalisation factor, and is 
derived by imposing that the cosmic rate of NSM (see Eq. ( \ref{eqn:rate})) reproduces the present one as determined by \citet{2021ApJ...913L...7A} ($\mathcal{R}=320^{+490}_{-240}$ Gpc$^{-3}$ yr$^{-1}$). \\
In general, \kalpha\ and \ansm\ could be time dependent variables. For example, the IMF could depend on time and the evolution of binary systems could be influenced by the metallicity. For the sake of simplicity, in this work both will be assumed constant in time.\\
Rather than proceeding directly to the evaluation of \ansm\ by solving Eq. (\ref{eqn:rate}) with the inclusion of a function $\psi(t)$ describing the cosmic SFH, as, e.g. in \citet{Simonetti2019MNRAS.486.2896S},  we proceed constructing a model for the galaxy population in the Universe which conforms to the major observational constraints. In this way, we are able to characterise the expected properties of the hosts of the NSM events.\\

\subsection{The sample of galaxies} \label{sec:TheSampleofGalaxies}
In this Section we describe our model for the galaxy sample which complies with the observed mass distribution \citepalias[][]{Peng2010ApJ...721..193P}, the observed relation between the Star Formation Rate and the mass of the parent galaxy \citep[][]{Renzini2015ApJ...801L..29R}, and the cosmic Star Formation rate density \citep[][]{MD2014}. An essential ingredient to construct this sample is the description of the SFH in each galaxy. We choose to adopt the observationally motivated log-normal shape \citepalias{G13} characterized by two parameters:
the logarithmic delay time ($t_0$) and the width of the log-normal function ($\tau$). Here we present some arguments supporting this type of formulation, while more details can be found in \citetalias[]{G13}. 
First, the evolution of the star formation rate density (SFRD) over the cosmic time shows a clear growth (at early times), a peak (around $t_{\rm U}\sim3$ Gyr), and a fall at later epochs \citep[e.g.][]{MD2014}. This rise and fall pattern is hard to describe with 
SFH laws that include only a declining trend (such as $\tau$ models), so that a second parameter, that acts as the "starting time" of the $\tau$ model, should be included to reproduce the time evolution of SFRD. Second, the log-normal function depends on two parameters and so it provides good flexibility to fit the age distribution in real galaxies. Thus we describe the SFH in galaxies as characterised by a tuple [$t_0,\tau$], and expressed as:
\begin{equation}\label{eqn:SFR}
    {\rm SFR}(t,t_0,\tau) = \frac{1}{t\sqrt{2\pi \tau^2}}\exp{\left( \frac{\left( \ln{t} - t_0  \right)^2   }{2\tau^2}     \right)    } \ \ .
\end{equation}
\noindent
We obtained the masses, redshifts and tuples ($t_0,\tau$) of the G13 sample of galaxies from L. Abramson (private comm.). These are 2094 galaxies with $0.03 \leq z \leq 0.11$, for which the parameters ($t_0 ,\tau ,M_{\rm Gal}$) were determined by fitting their Spectral Energy Distribution. In Fig. \ref{fig:SFH} we plot a selection of the SFH laws of galaxies contained in the sample of \citetalias{G13}. A large variety of SFHs is represented, with some galaxies having basically only one short initial episode of star formation; others, with wider age distributions, and peaking at various cosmic times, up to very recent epochs.\\
Eq.(\ref{eqn:SFR}) is normalized to 1 for $t$ ranging from 0 to infinity. 
To obtain the value of the SFR in units of $\sunmass$ yr$^{-1}$ we need to apply a conversion factor so that the mass of a galaxy at cosmic time $t$ is given by:
\begin{equation}
M_{\rm Gal}(t)  = K_{\rm Gal} \int_0^t {\rm SFR}(t,t_0,\tau) dt \ \ .
\label {eq:Mgal}
\end{equation}
For each galaxy of the G13 sample we determined $K_{\rm Gal}$ by applying Eq. (\ref{eq:Mgal}) with the cosmic epoch $t$ related to the galaxy redshift $z$ through our adopted cosmological model.\\
The \citetalias{G13} sample includes galaxies with masses greater than $10^{10}$ $\sunmass$, and lacks low mass galaxies, e.g. with masses ($10^8 \leq M_{\rm Gal} \leq 10^{10}$) $\sunmass$. Although the contribution to the cosmic SFRD from this low mass component is likely small, due to their low SFR, their number is much larger than the one of massive galaxies \citepalias[see][]{Peng2010ApJ...721..193P} and their total contribution to the NSM events may be important.
Furthermore, we point out that $\sim 1/3$ of SGRBs lack a coincident host galaxy \citep[see][]{Berger2014ARA&A..52...43B,Oconnor2022arXiv220409059O}. For example, \citet{Oconnor2022arXiv220409059O} found that the $28\%$ of the SGRBs that they have analysed are host-less. One possible explanation for this is that, at least in a fraction of cases, the hosts are faint galaxies at high redshift. \\
In order to model the NSM events in the Universe we thus need to add the low-mass galaxies component to the \citetalias{G13} sample. To do that we consider the galaxy stellar mass function (GSMF) in \citetalias{Peng2010ApJ...721..193P} which was determined adopting a Chabrier IMF. Since, the masses of the \citetalias{G13} galaxies where determined using a Salpeter IMF we apply a correction factor of 1/0.65 to the \citetalias{Peng2010ApJ...721..193P} masses, to re-scale them to the Salpeter IMF. It turns out that the GSMF of the G13 sample is overpopulated in the range $M_{\rm Gal} >10^{10} \: \sunmass$ with respect to the one in \citetalias{Peng2010ApJ...721..193P}. This tension has been noted also by \citet{Abramson2016ApJ...832....7A}. Possibly, the \citetalias{G13} sample targets a relatively high-density region in the Universe, compared to the general field. A more detailed discussion on this topic is reported in Appendix \ref{app:Overdensity}. Therefore, we adjust the high mass component in the \citetalias{G13} sample by lowering the number of galaxies by a factor of $0.85$ in all mass bins with $M_{\rm Gal} > 10^{10}$ $\sunmass$, and match the modified mass distribution to the \citetalias[]{Peng2010ApJ...721..193P} GSMF at $M_{\rm Gal} = 10^{10}$ $\sunmass$.\\
Concerning the functional form for the SFH in these low-mass galaxies we maintain the log-normal shape, fixing the $(t_0, \tau)$ tuples in such a way that the relation between the current SFR and the mass of the galaxy by 
\citet{Renzini2015ApJ...801L..29R} is fulfilled. More in detail, we first perform a random extraction for a galaxy mass in the range $10^8 \leq M_{\rm Gal}<10^{10}$ $\sunmass$ following the distribution in Fig. \ref{fig:MDF}. Then to each new synthetic galaxy we associate a current SFR given by:
\begin{figure*}
    \centering
    \includegraphics[trim={0cm 0cm 0cm 0cm}, clip,width=0.8\textwidth]{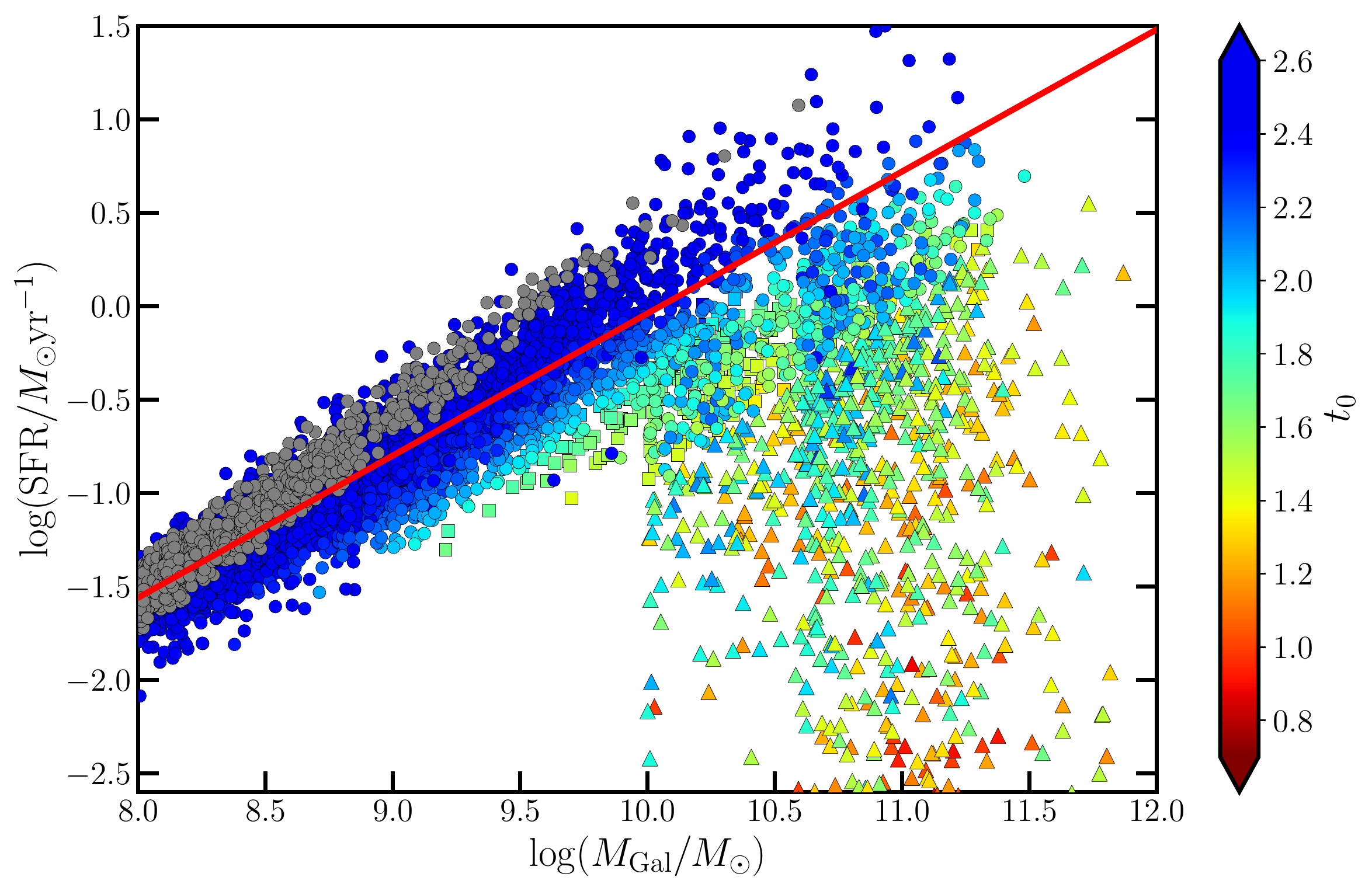}
    \caption{SFR vs $M_{\rm Gal}$ of the galaxies in our sample. The red solid line shows the star forming main sequence derived by \citet{Renzini2015ApJ...801L..29R}. Galaxies are labelled as reported in Figure \ref{fig:colorCode}. }
    \label{fig:SFRvsMass}
\end{figure*}
\begin{equation}\label{eq:5}
    {\rm SFR}_i^{\rm obs} = {\rm SFR}_{\rm MS} \times \mathcal{N}(1, \sigma^2(M_{{\rm Gal},i})) 
\end{equation}
where ${\rm SFR}_{\rm MS}$ is the Star Forming Main Sequence (SFMS) of galaxies in \citet{Renzini2015ApJ...801L..29R}:
\begin{equation}\label{eqn:MSRenzini}
    \log({\rm SFR}_{\rm MS}) = 0.76 \log(M_{\rm Gal}/\sunmass) - 7.64
\end{equation}
and $\mathcal{N}$($\mu$,$\sigma^2$) is a normal distribution with $\mu=1$ and  $\sigma^2$ is function of the stellar mass of the $i$-th galaxy. The inclusion of this second term allows us to describe the width of the galaxies distribution around the mean locus of Eq. (\ref{eqn:MSRenzini}).
Thus the tuple [$t_0,\tau$] must satisfy the following conditions:
\begin{equation}\label{eqn:BCMass}
    0.7 \times \int_0^{t_{\rm obs}}{\rm SFR}_i(t,t_0,\tau) = M_{{\rm Gal},i}^{\rm obs}
\end{equation}
and
\begin{equation}\label{eqn:BCSFR}
    {\rm SFR}_i(t_{\rm obs},t_0,\tau) = {\rm SFR}_i^{\rm obs} \ \ .
\end{equation}
Note that we apply a factor of 0.7 to account for the mass return of the stellar populations for a Salpeter IMF. In general this factor mildly depends on the SFH and age  \citep[see][]{GreggioBook2011spug.book.....G}.\\
Using Eqs from \ref{eq:5} to \ref{eqn:BCSFR}, for each synthetic galaxy we thus have the sSFR at z=0. This corresponds to a locus on the $t_0$-$\tau$ plane (see \citetalias{G13} Fig. 2) because of the degeneracy between the two parameters. In other words, the same value of sSFR can be obtained with an array of ($t_0$,$\tau$) tuples. We extract randomly one of these tuples having pre-computed the sSFR on an uniform $t_0$-$\tau$ grid. We checked that the stochasticity introduced by this random process has a negligible impact on our results.\\
The process is repeated until the sample contains the adequate number of galaxies to reproduce the GSMF by \citetalias{Peng2010ApJ...721..193P}. 
Proceeding in this way, we added 15850 galaxies with $8\leq \log(M_{\rm Gal}/\sunmass) <10$ to the 2094 original ones of \citetalias{G13}. In Fig. \ref{fig:MDF} we plot the GSMF of our final sample of 17944 galaxies. We notice that our mock catalogue still contains some excess of massive galaxies with respect to the P10 distribution, in spite of the 0.85 reduction of the original distribution in the Abramson sample. This is however the best compromise we found to match the two observational galaxy samples.
\subsubsection{Properties of the new sample}
\begin{figure}
    \centering
    \includegraphics[trim={0cm 0cm 0cm 0cm}, clip,width=0.5\textwidth]{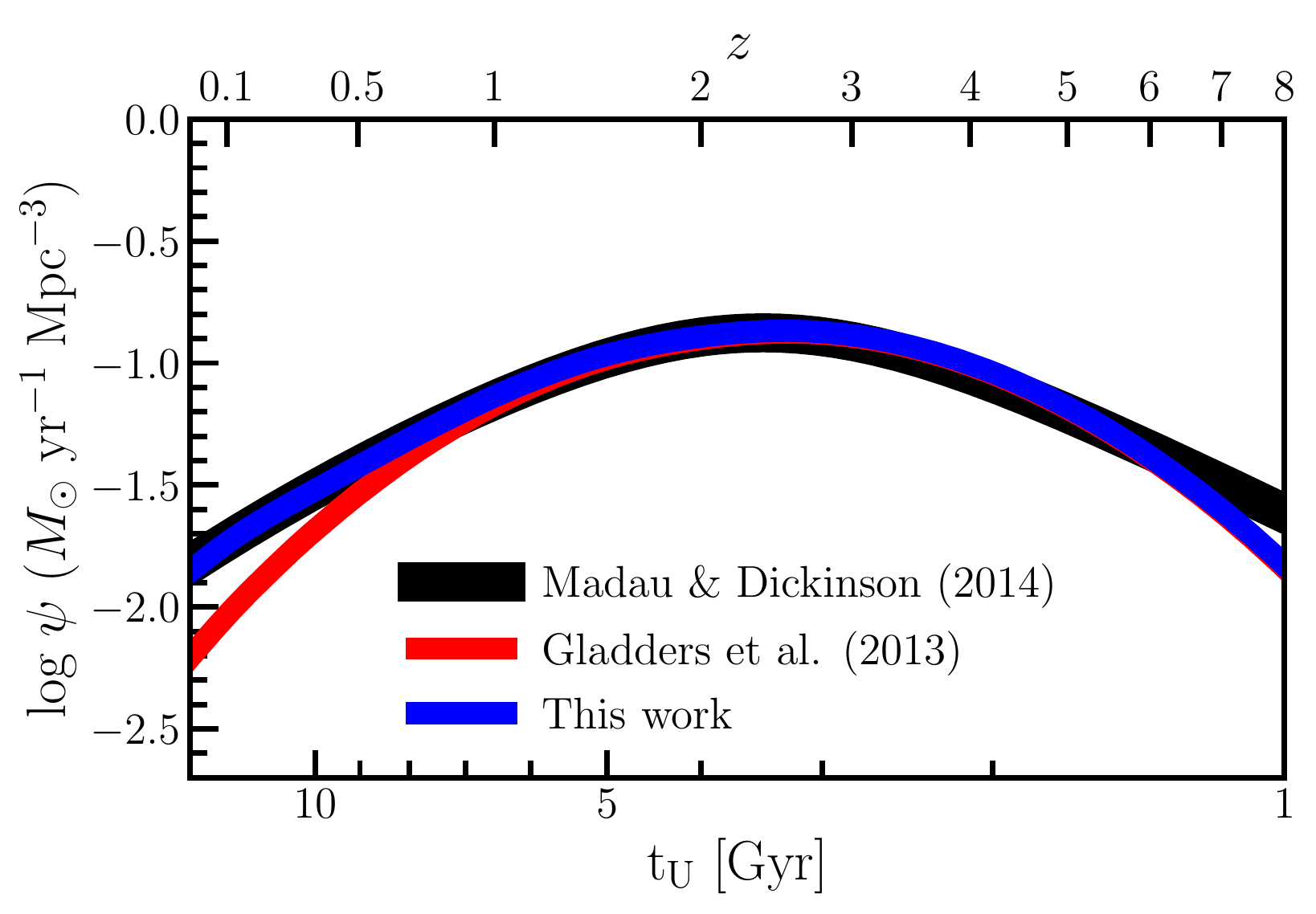}
    \caption{Cosmic SFRD vs redshift. Comparison between the cosmic SFRDs obtained from: \citet{MD2014} (black), \citetalias{G13} sample (red), and the mock sample developped in this work (blue).} 
    \label{fig:SFRD}
\end{figure}

In Fig. \ref{fig:colorCode} we plot the $(t_0,\tau)$ tuples of the galaxies
which compose our sample labelled as follows:\\\\
\textit{Color}: encodes to the value of $t_0$. This parameter characterises the time at which the SFR peaks. In Fig. \ref{fig:colorCode} we see that in the $\tau$ vs $t_0$ plane galaxies of our sample are distributed in a triangular-like pattern. This means that, galaxies that peak at early times ($t_0<1$), labelled in red, have SFHs with narrow peaks (small $\tau$)
(see also Fig. \ref{fig:SFH}). Some galaxies of our extended sample have a very late peak, at $z\gtrsim0$. They are plotted with grey dots.\\\\
\textit{Shape}: galaxies are also divided into three different groups based on both the sSFR and the epoch at which the SFR peaks: i) galaxies with low sSFR at $z\sim0$ ($<10^{-11}$ yr$^{-1}$) are plotted with triangles. Galaxies with low sSFR are generally called quiescent; ii) early-peak galaxies ($z_{\rm peak}>2$) with high sSFR at $z\sim0$ ($>10^{-11}$ yr$^{-1}$) are plotted with squares; iii) galaxies with high sSFR that peaks at $z_{\rm peak}<2$ are plotted with circles.\\\\
As a consistency check, we plot in Fig. \ref{fig:SFRvsMass} the SFR as a function of the stellar mass at $z\sim0$ for the galaxies of our sample to which we superimpose the SFMS relation of \citet{Renzini2015ApJ...801L..29R}. \\
In the low mass range $(8\leq \log(M_{\rm Gal}/\sunmass) \leq 10)$ the properties of our mock galaxies are fully consistent with the SFMS of \citet{Renzini2015ApJ...801L..29R} by construction. 
However, our sample lacks galaxies with mass around $10^9$ \sunmass\ and very low SFR ($\sim 0.01$ \sunmass yr$^{-1}$) which are present in the \citet{Renzini2015ApJ...801L..29R} sample.\\
Since our mock sample fulfils the GSMF this galaxy population is present among the \starf\ low-mass objects. Therefore, our models will overestimate the NSM events in low-mass galaxies at $z=0$. The discrepancy however is small since the contribution of low-mass galaxy to the NSM overall statistics in negligible (see Section \ref{sec:results}).\\
The sharp cut at $\log (M_{\rm Gal}/\sunmass)=10$ in Fig. \ref{fig:SFRvsMass} is generated by the method used to build our sample, in that we assumed that all the new synthetic galaxies (with $10^8 \leq M_{\rm Gal} \leq 10^{10}\ \sunmass$) added to the \citetalias{G13} sample, follow the SFMS.\\
We acknowledge that the log-normal description of the SFR results in an extended distribution of objects with a sSFR $< 10^{-11}$ yr$^{-1}$ while \citet{Renzini2015ApJ...801L..29R} shows a clump of \passive\ galaxies. On the one hand, this can be a limit of the functional form adopted to describe the SFH in our galaxies; on the other hand, the observational estimate of the sSFR could become inaccurate below a certain threshold (see discussion in \citetalias[]{G13}). As a result on Fig. \ref{fig:SFRvsMass} the region of \passive\ galaxies is poorly populated, compared to the one showed in Fig. 4 of \citet[]{Renzini2015ApJ...801L..29R}. However, focusing on galaxies with $10\leq \log(M_{\rm Gal}/\sunmass) \leq 11$, our mock catalogue contains $\sim 700$, $200$ and $600$ objects respectively with sSFR$>10^{-11}$ yr$^{-1}$, in $10^{-11.5}<$sSFR$<10^{-11}$ yr$^{-1}$ and lower than $10^{-11.5}$ yr$^{-1}$. These proportions are not dissimilar from those in Fig. 4 of \citet[]{Renzini2015ApJ...801L..29R}.\\\\
\noindent
The third constraint that our mock sample needs to satisfy is the evolution of the SFRD of \citet{MD2014}.\\ The evolution of the SFRD predicted by our sample of galaxies is 
\begin{equation}
    {\rm SFRD}(t) = \frac{1}{V}\sum_{[t_0,\tau]_i} {\rm SFR}_i(t,t_0,\tau)
    \label{eqn:SFRD}
\end{equation}
where $V$ is the volume that contains our galaxies. The volume sampled by a search over an area of $\Theta$ deg$^2$ spanning the redshift range from $z_1$ to $z_2$ is:
\begin{equation*}
    V = \frac{4\pi}{3}\frac{\Theta}{41253} \left[ \frac{c}{\rm H_0} \int_{z_1}^{z_2} \frac{dz'}{\sqrt{\Omega_{\rm M}(1+z')^3+\Omega_{\Lambda}}}  \right]^3 
\end{equation*}
where $\rm H_0$, $\Omega_{\rm M}$, and $\Omega_{\Lambda}$ depend on the assumed cosmological model. The redshift range of the galaxies in the sample of \citetalias{G13} is from $z_1=0.03$ to $z_2=0.11$ while the survey area is $\Theta=86$ deg$^2$. Thus we obtain a cosmic volume of:
\begin{equation}
    V = \num{3.18e5} \text{ Mpc}^3 \ \ .
\end{equation}
In Fig. \ref{fig:SFRD} we compare the evolution of SFRD of \citet{MD2014} (black) to that resulting from Eq. (\ref{eqn:SFRD}) when summing on the original \citetalias{G13} sample (red), and on the total galaxy sample developed in this work (blue). It appears that the original sample of \citetalias{G13} lacks star-forming galaxies in the local Universe ($z<1$). However, with the addition of the low-mass (and star-forming) population of galaxies, our sample provides a very good agreement with the SFRD found by \citet{MD2014}. As discussed in Section \ref{sec:TheSampleofGalaxies}, our sample presents a clear over-density with respect to the GSMF of \citetalias{Peng2010ApJ...721..193P}. In spite of this, and in spite of the differences between the distribution of galaxies on the (SFR, Mass) plane in \citet{Renzini2015ApJ...801L..29R} and ours noticed above, the description of the cosmic SFH in our model is very good.\\
We conclude that our mock galaxy sample satisfies the major empirical relations: the GSMF of galaxies, the relation between the current SFR and galaxy mass, and the cosmic SFRD as a function of redshift.\\
\begin{figure*}
    \centering
    \includegraphics[width=0.93\textwidth]{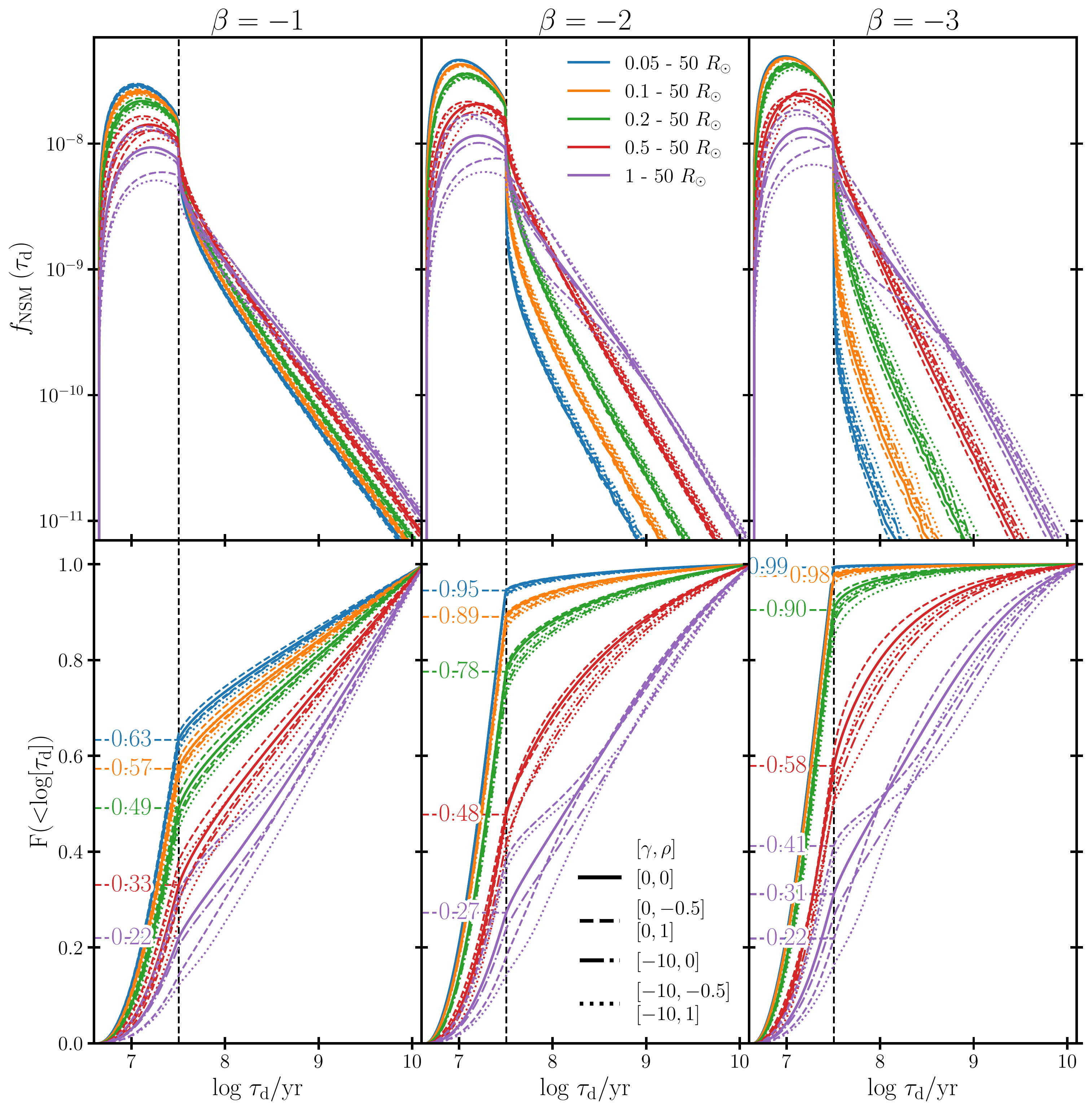}
    \caption{Differential and cumulative distributions of the delay times (upper and bottom panels, respectively) for models with $\beta=-1$ (left panels), $\beta=-2$ (central panels), and $\beta=-3$ (right panels). 
    Models obtained with different values for the minimum separation are plotted with different colours, according to the encoding labelled in the top central panel. Models with flat distributions of eccentricities and total mass (i.e. $[\gamma,\rho] = [0,0]$) are plotted with solid lines. Dashed lines show models adopting a flat distribution of the total binary masses and different distributions of the eccentricities (i.e. $[\gamma,\rho] = [0,-0.5]$ and $[0,1]$). Models computed with a steep distribution of the binary masses are plotted with dotted lines (i.e. $[\gamma,\rho] = [-10,-0.5]$ and $[-10,1]$). The dot-dashed lines show the results for the combination $[\gamma,\rho]=[-10,0]$. Vertical dotted lines are plotted at $32$ Myr, i.e. the lifetime of a $9$ $\sunmass$ star. In the bottom panels, we also report the fraction of prompt systems.}
    \label{fig:DTDs}
\end{figure*}
\subsection{The delay time distribution function}\label{sec:DTD}
As previously mentioned the DTD function is the second crucial ingredient needed to calculate $R_{\rm NSM}$ (see Eq. (\ref{eqn:rate})). In this work, we use the DTDs of NSM developed by \citet{Greggio2021}. For a NSM, the time between the formation of the binary system of stars and the merging event is equal to the time it takes to the secondary component to evolve into a neutron star (hereafter referred to as nuclear timescale), plus the time in which the two neutron stars are brought into contact due to the emission of gravitational waves (hereafter referred to as gravitational delay). The nuclear timescale ($\tau_n$) is a function of the initial mass of the star and its chemical composition. Neglecting the dependence on the chemical composition \citet{Greggio2021} adopted the following relation:
\begin{equation}
    \log m = 0.49 \left( \tau_n \right)^2 - 7.80 \log \tau_n + 31.88
\end{equation}
where $m$ and $\tau_n$ are in units of $\sunmass$ and yr, respectively. \\
The gravitational delay can be expressed by the following approximate relation (see \cite{Greggio2021}):
\begin{equation}\label{eqn:tauGW}
    \tau_{\rm GW}= \frac{0.6A^4}{M_{\rm DN}^3}\times(1-e^2)^{7/2} \text{ Gyr}
\end{equation}
where $A$, $M_{\rm DN}$, and $e$ are respectively the separation, the total mass (both expressed in solar units), and the eccentricity of the binary neutron star system at birth. 
Assuming that these three parameters follow power law distributions characterized respectively by the exponents $\beta$, $\gamma$ and $\rho$, \citet{Greggio2021} computed the distribution of the gravitational delays via Monte-Carlo realizations. It is worth noticing that these simulations include a correlation between the eccentricity of the binary system when the second neutron star is born and the separation $A$, mimicking the effect of the Supernova kick as suggested in the models by \citet{andrews_and_zezas_19}. The distribution of the gravitational delays convolved with the distribution of nuclear delays, resulted in parametrized DTDs. 
The assumption of a power law distributions for the parameters $A, M_{\rm DN}$ and $\rho$ is hinted by the results of Binary Population Synthesis computations (see \citet{Greggio2021} for details). \\
In addition to the slope of the power-law distribution of the separations, the DTD also depends on the minimum separation assumed. \citet{Greggio2021} explored the possibility that the minimum separation $A_{\rm min}$ can assume the following values: $0.05, 0.1, 0.2, 0.5, 1$ $R_{\odot}$. Although very low values of $A_{\rm min}$ are not found in numerical realisations of BPS, it is nonetheless worth examining these possibilities, since they provide a large number of events with short gravitational delays and this parameter may prove important for the merging rate in star-forming galaxies.\\
The general shape of these DTDs is as follows. The minimum delay time of all DTDs is $\tau_{\rm d} \sim 4.5$ Myr, that is the nuclear evolutionary lifetime of the most massive neutron star progenitor. At short delay times, DTDs are characterised by a wide peak ranging from $4.5$ Myr to $\simeq 32$ Myr,
the latter being the nuclear evolutionary timescale of the least massive NS progenitor. 
At delay times longer than 100 Myr, these DTDs can be well described by a power-law with a slope $s=0.25\times\beta-0.75$.
In the following we will tag as \textit{prompt} those events with a delay time shorter than $\tau_{\rm d}=32$ Myr and \textit{delayed} those with longer delay times.\\
In Fig. \ref{fig:DTDs} we plot the differential (top) and cumulative (bottom)  distributions of the total delay times of NSMs, for different separation ranges (drawn with different colours), values of $\beta$ (from the left to the right panels), and tuples of [$\gamma,\rho$] (plotted with different line-styles).
The values adopted for these parameters are meant to bracket a wide range of possibilities, i.e. a distribution of the binary masses skewed at the lower mass end ($\gamma=-10$) or flat ($\gamma=0$), as well as a variety of shapes for the distribution of the eccentricities.\\
Looking at the cumulative distributions (bottom panels of Fig. \ref{fig:DTDs}), it is possible to appreciate the dependence of the DTDs from the $\beta$ parameter. In particular, models with smaller values of $\beta$ predict, on average, DTDs which are more populated at the short delay times. At fixed $\beta$, models with wider separations (i.e. $1-50$ R$_{\odot}$) are more populated at long delay times.
This second trend is driven by the relatively large value of A$_{\rm min}$ and we remark that the dependence on the maximum separation is very mild \citep[see][]{Greggio2021}. There is some dependence of the DTD on the distributions of the binary neutron star total masses and on the eccentricities, as shown by the broken lines. However, it appears that the DTD is mostly sensitive to variations of the parameters $\beta$ and $A_{\rm min}$.\\
In the bottom panels of Fig \ref{fig:DTDs} we label the curves with different $A_{\rm min}$ by the fraction of systems that merge within 32 Myr since the episode of star formation, or the fraction of \prompt\ events.
At fixed $\beta$, models that assume closer systems (i.e. smaller $A_{\rm min}$) show larger fractions of \textit{prompt} events. A similar trend is obtained at fixed $A_{\rm min}$, when a steeper distribution of the separations is assumed (i.e. lower $\beta$).
We notice that similar fractions of \textit{prompt} events can be obtained with different combinations (e.g. $\beta=-1$; $A_{\rm min}=0.1$ $R_{\odot}$ and $\beta=-3$; $A_{\rm min}=0.5$ $R_{\odot}$), and that a very high fraction of \prompt\ events ($\gtrsim 90\% $) can be obtained for $\beta \lesssim -2$ and $A_{\rm min} \lesssim 0.2$. \\
The dependence of the DTD on the $\rho$ and $\gamma$ parameters is enhanced as the value of $A_{\rm min}$ increases. This is well visible in Fig. \ref{fig:DTDs} looking at the purple lines (A$_{\rm min}$ = 1 $R_\odot$). In particular, the DTDs show a strong dependence on the assumed distribution of the orbit eccentricities. For example, for $A_{\rm min}$ = 1 $R_\odot$ and $\beta=-3$, the fraction of \textit{prompt} events varies from $0.22$ to $0.41$ when $\rho$ changes from $-0.5$ to $0.5$, i.e. increasing the fraction of systems with high eccentricities (see right bottom panel of Fig. \ref{fig:DTDs}.\\
The difference between the fractions of \textit{prompt} and \textit{delayed} events has a great impact on the chemical enrichment caused by the NSM systems. As mentioned before, chemical evolution models suggest that the enrichment of r-process elements\footnote{elements mostly produced by the r-process channel} (such as Eu) should occur on short delay times. This can be accomplished either with a steep distribution of the separations and small $A_{\rm min}$, if Eu is produced only by NSMs, or with flatter DTDs for the NSM, if some contribution to Eu comes from massive stars.

\begin{figure*}
    \centering
    \includegraphics[trim={0cm 0cm 0cm 0cm}, clip, width=0.98\textwidth]{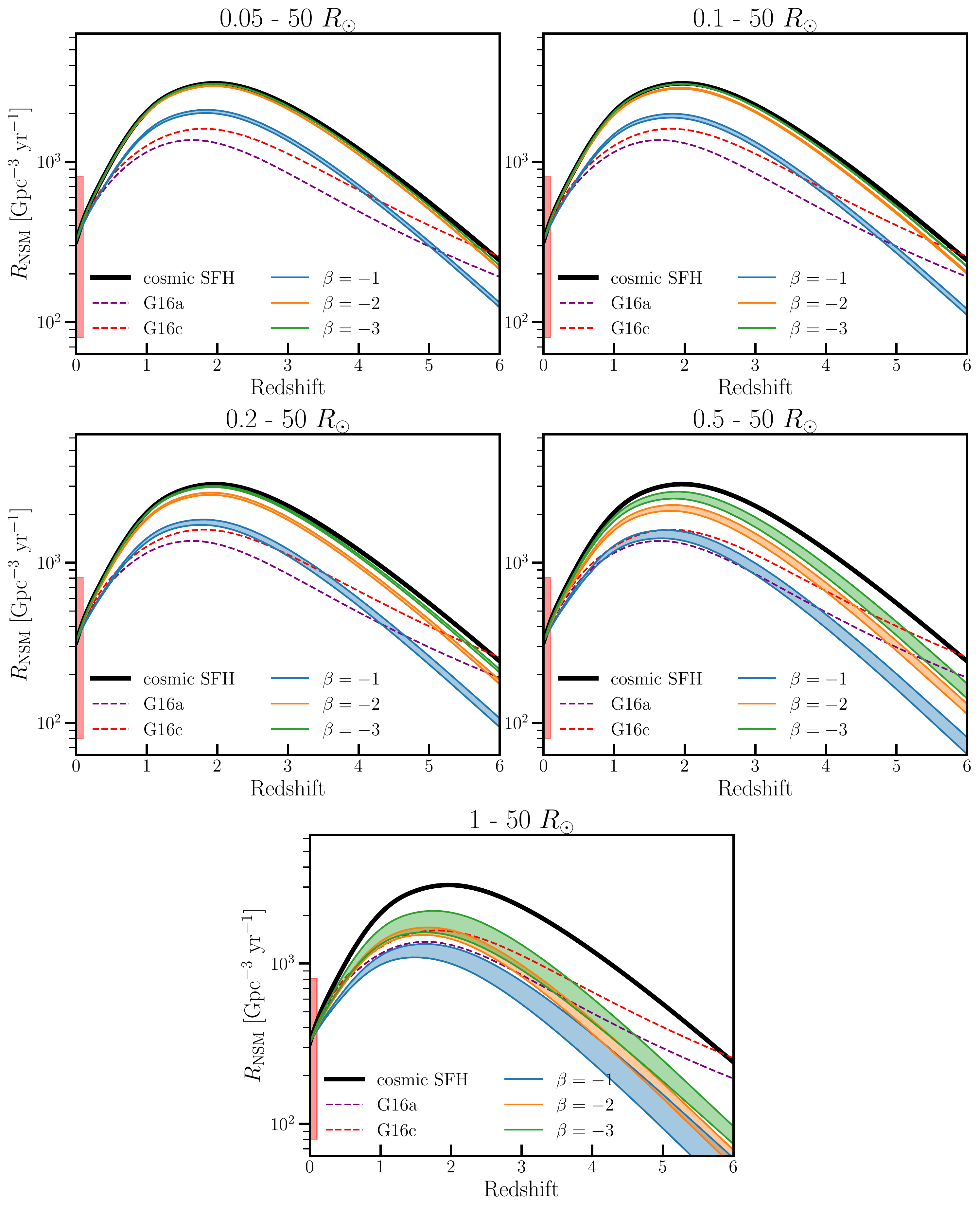}
    \caption{Redshift distribution of the predicted cosmic rate of NSMs for DTDs with different ranges for the initial separations and various values of $\beta$, as indicated. For each combination of the (A$_{\rm min}, \beta$) parameters, the shaded regions include all the models obtained with the range of the $(\gamma,\rho)$ parameters explored here. Models' predictions are compared with the redshift distribution of SGRB obtained by \citet{Ghirlanda2016} (red and purple dashed lines). The red vertical bar at zero redshift indicates the current estimate of the local rate of NSMs and its uncertainty ($\mathcal{R}=320_{-240}^{+490}$ Gpc$^{-3}$ yr$^{-1}$)  \citet{2021ApJ...913L...7A}. With a black solid line, we plot the trend of the cosmic SFH of our mock galaxy population.
    } 
    \label{fig:CosmicRate1}
\end{figure*}

\section{Redshift distribution of neutron star mergers} \label{sec:zresults}
In this section, we present the redshift distribution of cosmic NSM rates predicted by different DTD models useful for the calibration of \ansm\ . For each model, the parameter \ansm\ is determined by imposing that the present rate of NSM predicted by our models is equal to the one derived by \citet{2021ApJ...913L...7A} (i.e. $\mathcal{R}= 320$ Gpc$^{-3}$ yr$^{-1}$).\\
In addition we compare an observationally constrained SGRB redshift distribution with the one computed with Eq. (\ref{eqn:rate}). As previously introduced, NSMs are thought to be the progenitors of SGRB and so these two astrophysical events should present similar redshift distributions. For this comparison, we use the redshift distribution of SGRBs derived by \citet{Ghirlanda2016}. These authors performed a Monte Carlo simulation for a population of SGRBs described by a given luminosity function $\Phi(L)$ and a redshift distribution $\Psi(z)$ parametrized as follows:
\begin{equation} \label{eqn:SGRB}
\Psi(z) = \frac{1+p_1z}{1+\left(z/z_p \right)^{p_2}} \ \ .
\end{equation}
The parameters of the redshift distribution are then derived by fitting the properties of the synthetic population to a set of observational constraints derived from Fermi and Swift observations of SGRBs \citep[for more details see][]{Ghirlanda2016}. With these procedure the authors propose two sets of parameters which provide equally good fits to the observations: $p_1=2.8$, $p_2=3.5$, $z_p=2.3$ ("model a"), or $p_1=3.1$, $p_2=3.6$, $z_p=2.5$ ("model c").\\
Fig. \ref{fig:CosmicRate1} shows our models for the redshift evolution of the cosmic rate of NSMs computed with different assumptions on the DTD. The shaded areas show the range spanned by the rate for all the [$\gamma,\rho$] combinations of the specific choice of the [$\beta,\rm A_{\rm min}$] tuple.
In the panels we also plot the curves of "model a" (red dashed line) and "model c" (purple dashed line) of \citet{Ghirlanda2016}. At $z=0$, the red shaded area indicates the current rate of NSM evaluated by \citet{2021ApJ...913L...7A} with its uncertainty.  The black solid line shows the trend of the redshift evolution of the cosmic SFR of our mock sample, scaled to the local NSM merger rate.\\
From Fig. \ref{fig:CosmicRate1} we can note some features of $R_{\rm NSM}(z)$ connected to the characteristics of the assumed DTD model:
\begin{list}{--}{} 
    
    \item models with smaller $A_{\rm min}$ and/or steeper DTDs (i.e. lower values of $\beta$) predict a faster drop of the rate of NSM between the peak (around $z\sim2$) and $z=0$ compared to models with higher proportion of late events. It follows that models with a larger fraction of \textit{prompt} events predict a stronger evolution of the NSM rate with redshift.\\
    
    \item models for the redshift distributions characterized by small $A_{\rm min}$ and/or lower values of $\beta$ evolve closer to the cosmic SFRD. This follows from the high fraction of \textit{prompt} events characterizing these models.\\
    
    \item the dependence of the rate on the distribution of total mass and eccentricity of the neutron star binaries is generally small, but becomes more important for the cases with larger fraction of late events. This is shown by the width of the shaded areas in Fig. \ref{fig:CosmicRate1}.\\
    
    \item the best fitting redshfit distribution of SGRBs by \cite{Ghirlanda2016} are relatively flat, indicating $\beta$ between $-1$ and $-2$, in combination with $A_{\rm min}$ between $0.5$ and $1$ $R_\odot$. We regard the discrepancy at $z \gtrsim 4$ as an unimportant issue, since the \citet{Ghirlanda2016} relations are poorly constrained at high redshift.
\end{list}

\begin{figure}
    \centering
    \includegraphics[trim={0cm 0cm 0cm 0cm}, clip, width=0.5\textwidth]{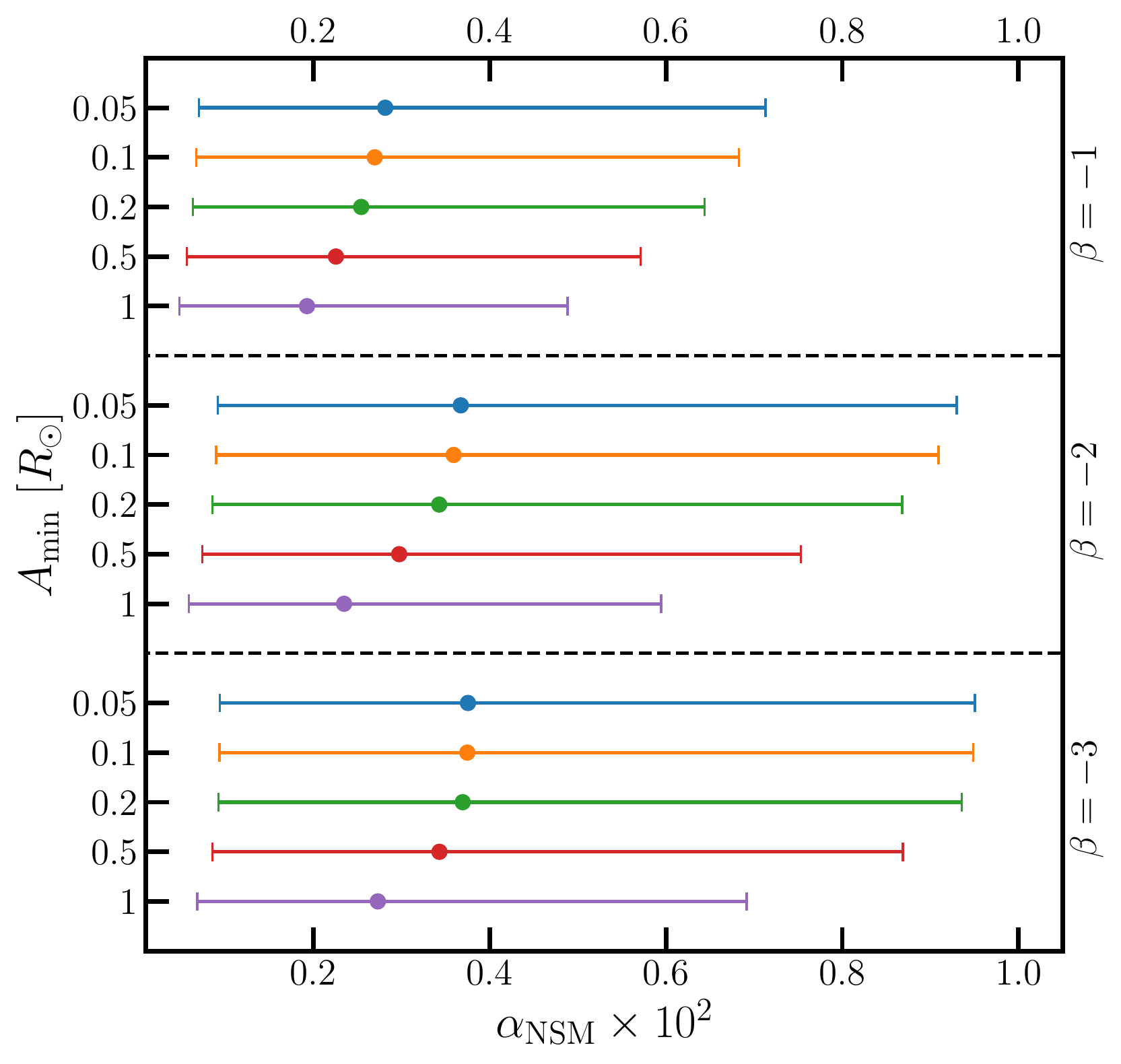}
    \caption{Values of \ansm\ with uncertainties for all the DTDs used in this work, calculated as described in Section \ref{sec:uncertain}. Values and errors are multiplied by $10^2$.}
    \label{fig:alphas}
\end{figure}
\begin{figure}
    \centering
    \includegraphics[trim={0cm 0cm 0cm 0cm}, clip, width=0.5\textwidth]{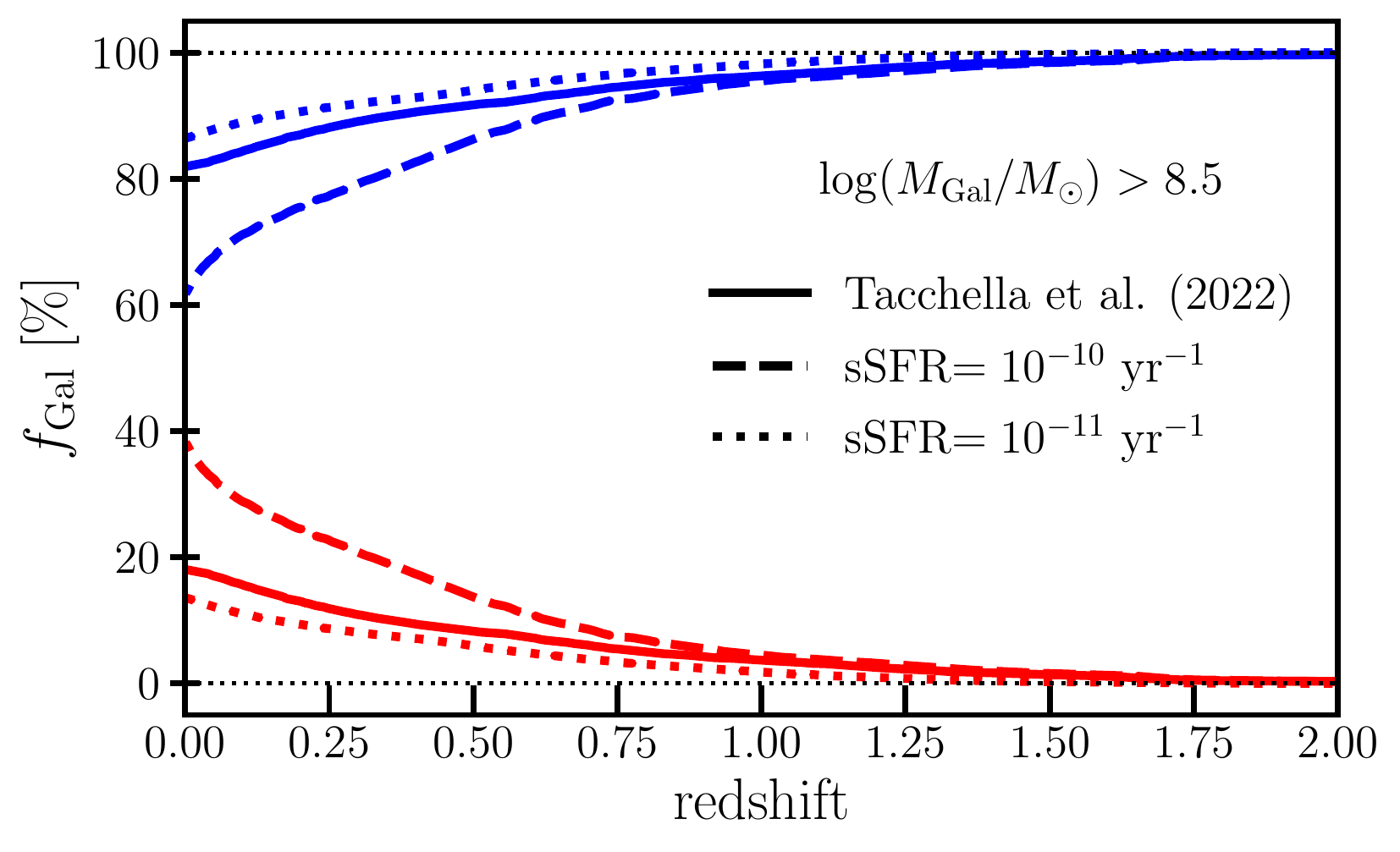}
    \caption{Redshift evolution of the fraction of \passive\ (red curves) and \starf\ (blue curves) galaxies (with $\log(M_{\rm Gal}/\sunmass)>8.5$) in our mock universe based on the chosen classification criteria.}
    \label{fig:fracGal}
\end{figure}
\begin{figure*}
    \centering
    \includegraphics[trim={0cm 0cm 0cm 0cm}, clip, width=0.9\textwidth]{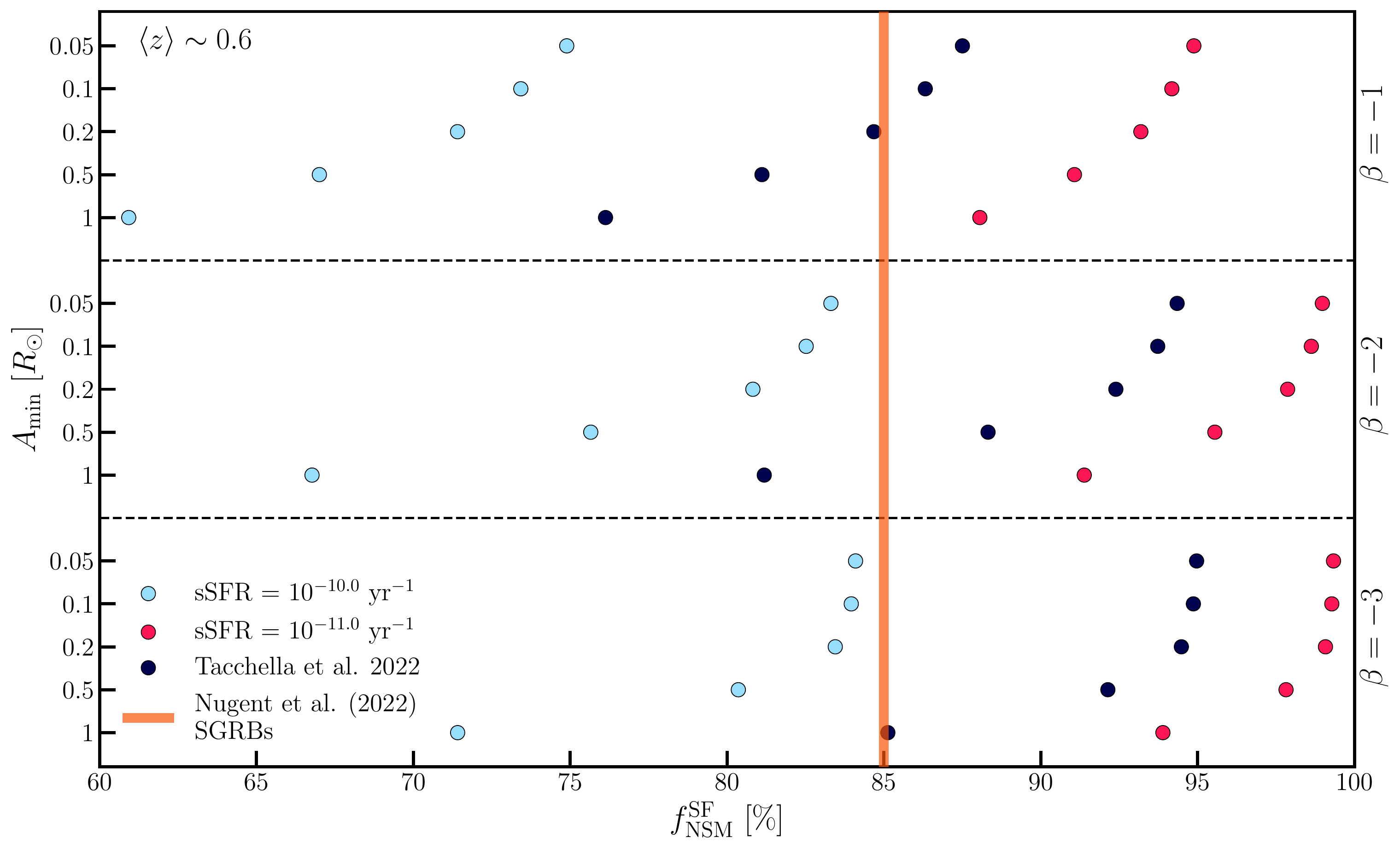}
    \caption{Predicted fractions of NSM in SF galaxies (at $z\sim0.6$) for different DTDs  (as labelled on the vertical axis), and different selection criteria (as labelled in the bottom panel). The orange vertical line shows the value ($85 \%$) retrieved for the population of SGRBs hosts analysed by \citet{Fong2022barXiv220601764N} to be compared to the black circles.} 
    \label{fig:fractions}
\end{figure*}
\subsection{Uncertainties on the scaling factor}\label{sec:uncertain}
As previously mentioned, we determine the factor \ansm\ by imposing that the theoretical curves in Fig. \ref{fig:CosmicRate1} reproduce the present rate of NSM of \citet{2021ApJ...913L...7A}. In our adopted approach, this factor is the fraction of massive stars that have the right characteristics to eventually lead to a NSM event within a Hubble time. Mathematically \ansm\ acts like a scaling factor of the rate of NSMs (see Eq. (\ref{eqn:rate})). The present-day rate of NSM presented by \citet{2021ApJ...913L...7A} has been estimated from the two NSM events detected so far observed (i.e. GW170817 and GW190425), so it presents a large uncertainty. As a consequence, also the estimated value of \ansm\ present uncertainties that arise directly from the ones of the observational rate. \citet{2021ApJ...913L...7A} report a most probable value of the rate and its $90 \%$ confidence levels of $\mathcal{R} = 320^{+490}_{-240}$ Gpc$^{-3}$ yr$^{-1}$. Correspondingly we report the most probable value of \ansm\ and its $90 \%$ confidence interval by imposing that our models reproduce the estimates by \citet{2021ApJ...913L...7A}. Fig. \ref{fig:alphas} shows our determinations of \ansm\ for all the DTDs used in this work. Assuming $\mathcal{R}=320$ Gpc$^{-3}$ yr$^{-1}$ we find that $\alpha_{\rm NSM}\sim 2-3\times 10^{-3}$. \\
From Fig. \ref{fig:alphas} we can note some correlations between the estimates of \ansm\ (and the uncertainties) and the assumed DTD model. Models with higher $A_{\rm min}$ and/or flatter DTDs predict the lower values of \ansm. This follows from the fact that, at given SFH, flatter DTDs produce more delayed events, so that at zero redshift the resulting cosmic NSM rate is larger than in the case of steeper DTDs.\\

\begin{figure*}
    \centering
    \includegraphics[trim={0cm 0cm 0cm 0cm}, clip, width=0.96\textwidth]{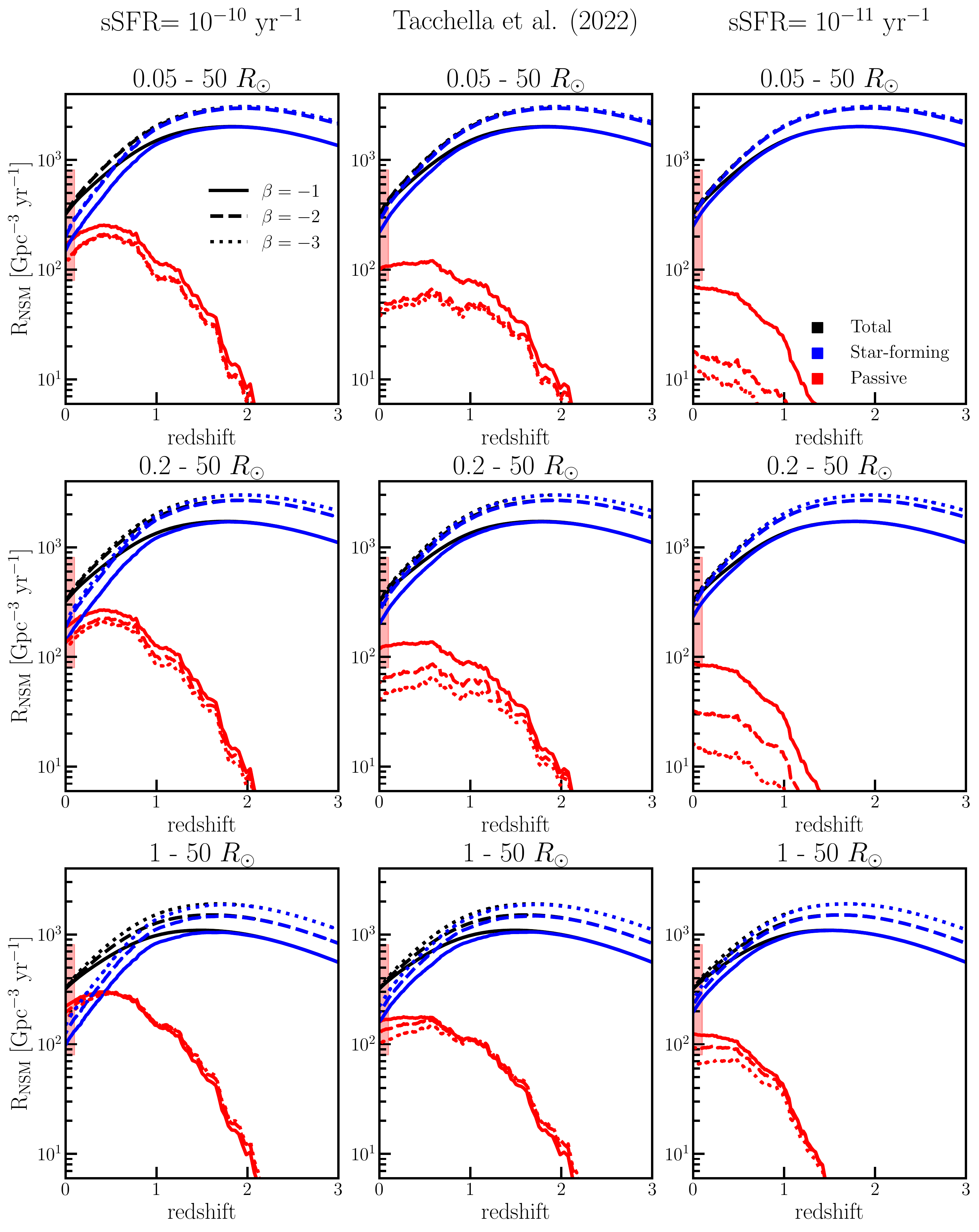}
    \caption{Cosmic rate of NSMs as a function of the redshift for \passive\ (red), SF (blue), and all (black) galaxies of our sample for different DTDs and for different criteria used to classify the galaxies, as labelled on top. Each panel reports the trends for the three values of $\beta$ (solid, dashed and dotted respectively for $\beta=-1,-2$ and $-3$), while the three rows show the results for $A_{\rm min}=(0.05,0.2$ and 1) $R_\odot$ from top to bottom.}
    \label{fig:fractionsvsz}
\end{figure*}
\begin{figure*}
    \centering
    \includegraphics[trim={0cm 0cm 0cm 0cm}, clip, width=\textwidth]{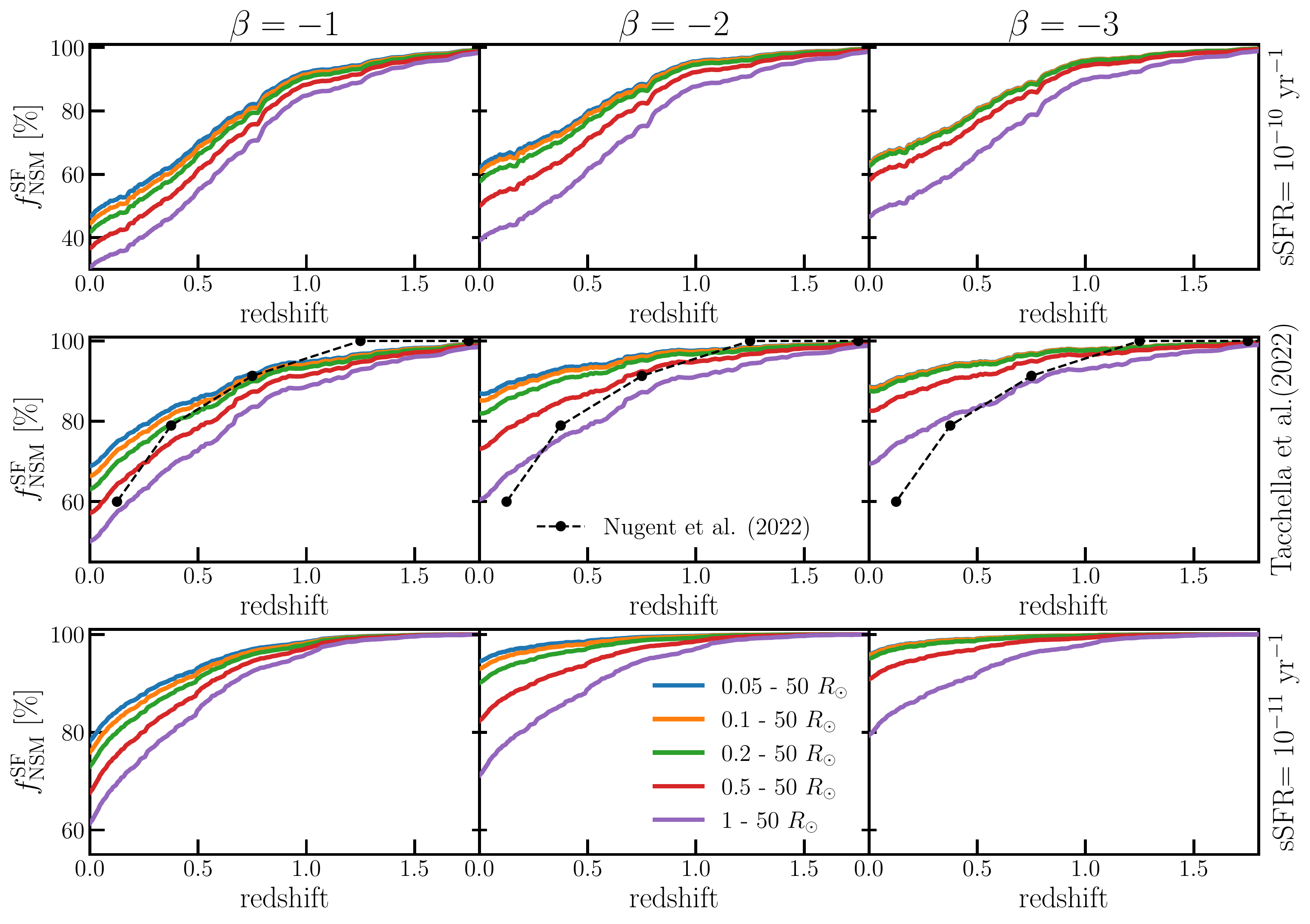}
    \caption{Redshift evolution of the fraction of NSMs hosted by SF galaxies for different DTDs. In the left, central, and right columns we plot the relations for DTD models with $\beta=-1$, $\beta=-2$, and $\beta=-3$, respectively, while in all panels the colour encoding refers to the value of $A_{\rm min}$ as labelled on the top right one. The different rows show the fractions computed adopting different criteria to define a \starf\ galaxy, as labelled on the right. In black, we also show the redshift evolution of the fraction of SGRBs hosted by SF galaxies from the data in \citet{Fong2022barXiv220601764N} (see appendix B).}
    \label{fig:fracvsz}
\end{figure*}
\begin{figure*}
    \centering
    \includegraphics[trim={0cm 0cm 0cm 0cm}, clip, width=\textwidth]{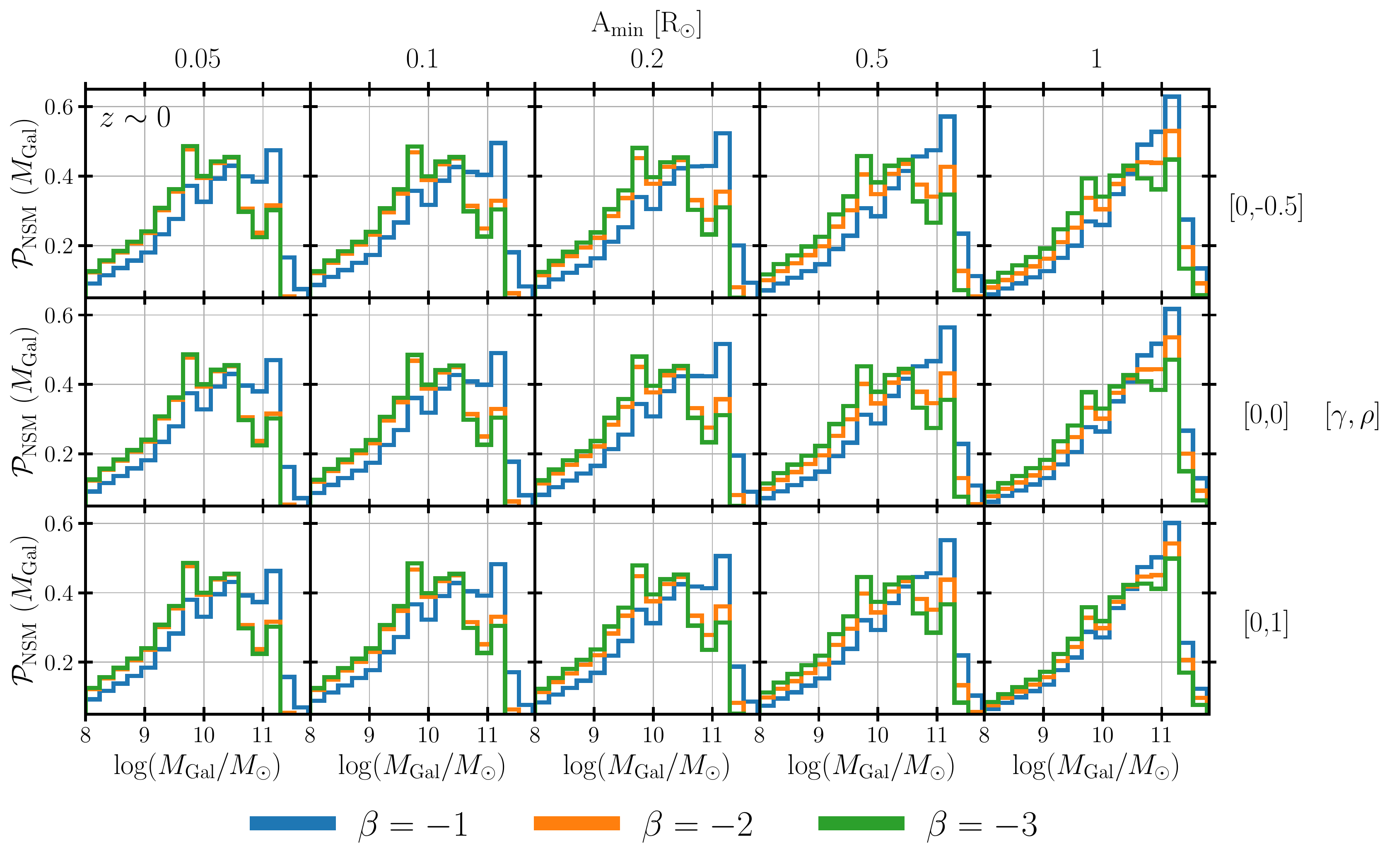}
    \caption{Distributions of $\mathcal{P}_{\rm NSM}(M_{\rm Gal})$ at $z=0$ for different DTDs with 30 different combinations of $A_{\rm min}$ and [$\gamma,\rho$]. For each combination we report $\mathcal{P}_{\rm NSM}(M_{\rm Gal})$ obtained with different $\beta$ in blue ($\beta=-1$), orange ($\beta=-2$), and green ($\beta=-3$). }
    \label{fig:Probabilityatz0}
\end{figure*}

\section{Gamma Ray Bursts and host galaxies} \label{sec:fractions}
GRBs are traditionally divided in two classes according to the duration of the burst: the short and the long GRBs that last respectively less or more than two seconds. Long GRBs are thought to be related to the death of massive stars for several observational reasons: (i) their association to Type Ic core-collapse SNe \citep[][]{Galama1998Natur.395..670G,Hjorth2003Natur.423..847H,Stanek2003ApJ...591L..17S,Woosley2006ARA&A..44..507W}; (ii) the fact that long GRBs are hosted only by star forming galaxies \citep[][]{Bloom1998ApJ...507L..25B,Djorgovski1998ApJ...508L..17D,Christensen2004A&A...425..913C,Wainwright2007ApJ...657..367W}, and (iii) their spatial coincidence with star forming regions \citep[][]{Bloom2002AJ....123.1111B,Fruchter2006Natur.441..463F}. These observational facts strongly support the notion that long GRBs are produced at the death of massive stars \citep[see][]{Roy2021Galax...9...79R}. 
On the other hand, based on their short duration, the progenitors of SGRBs are thought to be systems characterised by short dynamical timescales (e.g. merging of compact objects) \citep[see][]{Ciolfi2018IJMPD..2742004C}.\\
The association of GRBs with their host galaxies can provide more insights on their progenitors. 
Indeed, different from the case of long GRBs, SGRBs occur also in galaxies with low (if any) current SF. This is further evidence in favour of a different nature of the progenitors of short and long GRBs.\\
The presence of SGRBs in both \starf\ and \passive\ galaxies indicates that their progenitors are characterised by a wide range of delay times, similar to Supernovae of Type Ia. Because of this property, we expect a correlation between the rate of SGRBs and the average age of the parent stellar population sensitive to the shape of the DTD of NSM.
A steep DTD, with a large fraction of \textit{prompt}\ events, will predict more events in galaxies with vigorous star formation compared to those occurring in passive ones. Similar to the case of Type Ia Supernovae, the rate of NSMs per unit mass of the parent galaxy will then decrease going from star-forming to passive galaxies with a gradient which depends on the actual shape of the DTD.\\
Early studies on the hosts of few observed SGRBs had concluded that the population of hosts is dominated by \passive\ galaxies and so, the progenitors of SGRBs were thought to be objects with long delay times \citep[see][]{Prochaska2006ApJ...642..989P,Nakar2007PhR...442..166N,Gal-Yam2008ApJ...686..408G}. More recent studies, based on a larger sample of SGRBs, have concluded that only $\sim20 \%$ of the population of host galaxies are \passive\, while the majority of events occur in \starf\ galaxies \citep[][]{Leibler2010ApJ...725.1202L,Fong2013ApJ...769...56F, Berger2014ARA&A..52...43B}.\\
Further information regarding the SGRBs progenitors comes from the recent realisation that events classified as long GRB may be connected to NSMs. The long GRB 211211A has been followed by a kilonova emission with a similar luminosity, duration, and colour of AT 2017gfo, the event associated with GW 170817, suggesting that NSMs could be the progenitors of a sub-population of long GRBs \citep{Rastinejad2022Natur.612..223R}. This hypothesis has been reinforced by the detection of the peculiar GRB 060614 \citep[][]{Gehrels2006Natur.444.1044G}.
Because of its duration of $\sim102$ s, the event can be classified as a long GRB; however, its temporal lag and peak luminosity are typical of the short GRBs group. The measured duration of a GRB depends on its luminosity: a long GRB could remain above the detection threshold for a short time if it was intrinsically less energetic and/or located at a larger distance. This calls into question the classification of events into long and short, which should take into account the possible dispersion of intrinsic luminosities and distances of the GRBs. In other words, GRB 060614 can be seen as a long GRB that could have been regrouped among the SGRBs if it were less energetic or observed at a larger distance \citep[see][]{Zang2007ApJ...655L..25Z}. Furthermore, \citet{Yang2022arXiv220412771Y} have reported a significant excess in the optical and in the near-infrared emission of GRB 060614. This can be explained as a kilonova emission. There are then indications that some long GRBs lacking of supernova association, could originate from a NSM.\\
Recently, \citet{Fong2022aarXiv220601763F} have presented a census of the 90 SGRBs observed from 2005 to 2021 that have an association with a host galaxy. In their sample they have included all the short GRBs discovered by the Swift Observatory \citep[][]{Gehrels2004ApJ...611.1005G} that have: i) a duration shorter than $2$ s; ii) events with a coincident afterglow within a $5\arcsec$ radius. Their sample also includes the two peculiar long GRBs just discussed (GBR 211211A and GRB 060614) and a third one, GRB 160303A \citep[][]{Ukwatta2016GCN.19148....1U}, a peculiar long GRB with a $\sim0.4$ s peak followed by a tail to $\sim 5$ s, albeit with a low signal to noise.\\
\citet{Fong2022barXiv220601764N} analysed the sample presented by \citet{Fong2022aarXiv220601763F}. In particular, they used spectroscopy and optical and near-infrared photometry to characterise the stellar population properties of the host galaxies of SGRBs. They found that $\sim 85 \%$ of the population of hosts are \starf\ galaxies, $\sim 6 \%$ are transitioning galaxies, and $\sim 9 \%$ are quiescent galaxies.\\
In this section, we compare the fraction of NSMs in galaxies of different types predicted by our model of Universe to the indications from the observations. First, we need to specify a criterion to classify galaxies of different types. To do that we consider the sSFR with two values for the threshold to be classified as \starf\ (i.e. $10^{-10}$ $\rm yr^{-1}$ and $10^{-11}$ $\rm yr^{-1}$), and the criterion proposed by \citet[]{Tacchella2022ApJ...926..134T}, and adopted by \citet[]{Fong2022barXiv220601764N} to construct their statistics. This criterion considers the quantity
\begin{equation}
    \mathcal{D}(z) = {\rm sSFR}(z)\times t_{\rm U}(z)
\end{equation}
where $t_{\rm U}(z)$ is the age of the Universe at the redshift $z$, and a galaxy is classified as \starf\ if $\mathcal{D}(z)>1/3$. The quantity $\mathcal{D}(z)$ is the number of times the stellar mass doubles within the age of the universe at redshift $z$ assuming a constant sSFR over the time $t_{\rm U}(z)$. \citet[]{Tacchella2022ApJ...926..134T} divide galaxies into three categories, based on the value of $\mathcal{D}(z)$: \starf\ if $\mathcal{D}(z)>1/3$, transitioning if $1/3<\mathcal{D}(z)<1/20$, and quiescent if $\mathcal{D}(z)<1/20$. On the other hand, the classification criteria based on the sSFR divide galaxies into two categories: \starf\ (if the sSFR is above the threshold) and \passive\ (in the other case). To ease the comparison between results obtained with different classification criteria we merge the transitioning and quiescent classes of \citet[]{Tacchella2022ApJ...926..134T} into the \passive\ category (i.e. those with $\mathcal{D}(z)<1/3)$.\\
It is important to emphasize that different criteria used to separate \starf\ (SF) from \passive\ galaxies correspond to different partitions of the population of evolving galaxies in the Universe. Fig. \ref{fig:fracGal} shows the evolution with redshift of the fraction of galaxies classified as SF (and its complement) when these three different criteria are adopted. When requiring a higher sSFR threshold to classify a galaxy as SF, less objects satisfy the criterion. The difference of the fraction of SF galaxies becomes larger and larger as the redshift decreases, and more and more galaxies exit the range of high sSFR needed to be classified as SF. The criterion adopted by \citet[]{Tacchella2022ApJ...926..134T} produces an evolution of the fraction of star-forming galaxies close to what obtained with the sSFR $> 10^{-11} \: {\rm yr}^{-1}$  criterion. For each of the three different criteria, we compute the fraction of NSMs hosted by SF galaxies ($f_{\rm NSM}^{\rm SF}$) as:
\begin{equation}
    f_{\rm NSM}^{\rm SF}(z)  = \frac{R_{\rm NSM}^{\rm SF}(z)}{R_{\rm NSM}(z)}
\end{equation}
where $R_{\rm NSM}^{\rm SF}(z)$ is the rate of NSMs in SF galaxies at redshift $z$ and $R_{\rm NSM}(z)$ is the rate of NSMs from all the galaxies in our sample (SF + \textit{passive}).\\

\subsection{Our theoretical demographics} \label{sec:results}
From Fig. \ref{fig:fractions} to Fig. \ref{fig:fracvsz} the rate is plotted for different $\beta$'s and $A_{\rm min}$'s but no more dependence on $\gamma$ and $\rho$. We decide to restrict the discussion to the [$\gamma$,$\rho$]$=$[0,0] cases since the dependence on these parameters appears to be a second-order effect. In Fig. \ref{fig:fractions} we plot the fractions of NSMs in SF galaxies at $z\sim 0.6$ predicted by different DTDs. The choice of the redshift value is motivated by that of the average redshift of the \citet{Fong2022barXiv220601764N} sample. Fig. \ref{fig:fractions} clearly shows that DTDs characterized by a larger fraction of prompt events (i.e. steeper $\beta$ and/or smaller $A_{\rm min}$) produce a higher fraction of events in SF galaxies. The trend is the same for the three classification criteria. On the other hand, the value of this fraction is very sensitive to the adopted criterion. Indeed, the less demanding threshold for a galaxy to be classified as \starf\ (i.e. sSFR $> 10^{-11} {\rm yr}^{-1}$) implies that a larger number of galaxies are classified as such, so that the fraction $f_{\rm NSM}^{\rm SF}$ increases (see also Fig. \ref{fig:fracGal}).\\ 
These theoretical fractions can be compared to the corresponding value determined by \citet{Fong2022barXiv220601764N} of 0.85. We notice that when adopting the more demanding criterion for the classification of galaxies as SF, a steep DTD is required to account for such a large value of events in the relatively few \starf\ galaxies. On the opposite side, when choosing the threshold of sSFR $>10^{-11} \rm{yr}^{-1}$, most galaxies fall in the SF category, and so the fraction $f^{\rm SF}_{\rm NSM}$ is very large irrespective of the  DTDs. In our mock Universe, the criterion adopted by \citet{Tacchella2022ApJ...926..134T} yields constraints on the DTD, in that the observed fraction of SGRBs in SF galaxies can be met either with a relatively flat $\beta$ and a small $A_{\rm min}$ or with a steep $\beta$ and a larger value of $A_{\rm min}$.\\
Looking at Fig. \ref{fig:fractions} it appears that the large fraction of SGRBs observed in SF galaxies, points towards DTDs with a substantial fraction of \textit{prompt}\ events.\\
In Fig. \ref{fig:fractionsvsz} we plot the redshift evolution of the rates of NSMs in both \passive\ and SF galaxies, for different DTDs and different methods of galaxies classification. The left column shows the model prediction in the case in which SF galaxies are those with sSFR $>10^{-10}$ yr$^{-1}$. In general, the rate in SF galaxies dominates the total NSM rate at high redshift, up to the local Universe, where the flatter DTDs (higher $\beta$ and/or higher $A_{\rm min}$ correspond to similar contributions to the total rate of the two galaxy types.
For A$_{\rm min}=1$, at redshift lower than $\sim 0.5$ the majority of events should occur in galaxies with sSFR < $10^{-10} \: \rm{yr}^{-1}$ irrespective of $\beta$.\\
The right column in Fig. \ref{fig:fractionsvsz} shows the results for the lowest sSFR threshold. With this cut, more galaxies are classified as SF, compared to the previous cases, and so the contribution of \passive\ galaxies to the total rate of NSMs is extremely low, irrespective of the DTD. \\
The central column shows the predictions when adopting the \citet{Tacchella2022ApJ...926..134T} criterion for galaxies classification. Also in this case the rate in SF galaxies dominates at all redshifts, except for the flattest DTDs, which produce a similar contribution to the events in the two galaxy kinds at very low redshift.\\
Fig. \ref{fig:fractionsvsz} shows that the redshift evolution of the NSMs hosts properties has the potential to discriminate among the various choices for the DTD parameters. However, the operational criterion used to classify galaxies as SF or \passive\ is critical, and has a great effect on the fraction of NSM hosted by galaxies of different types.\\
In Fig. \ref{fig:fracvsz} we plot the redshift evolution of the fraction of NSMs hosted by SF galaxies for different DTDs, which can be compared to an observational counterpart derived from the data in \citet{Fong2022barXiv220601764N} (black circles). 
The observed fractions are affected by a large uncertainty (that covers the entire area of the plots in Fig. \ref{fig:fracvsz}) due to the shot noise caused by the small number of SGRBs observed in various redshift bins (see Appendix \ref{app:SGRBstatistic}). As a consequence, all the models fall into the error intervals of SGRBs observations of \citet{Fong2022barXiv220601764N}. So, with the current uncertainties of the observed redshift evolution of the fraction of SGRBs hosted by SF galaxies, no firm conclusion could be drawn. However, Fig. \ref{fig:fractionsvsz} shows that this kind of statistic has the potential to indicate the steepness of the DTD, especially in the nearby Universe. At redshifts larger than $\sim 1$ almost all events should instead occur in star-forming galaxies, which hampers the possibility of discriminating the DTD parameters. \\
In Appendix \ref{app:SGRBstatistic} we show that a $\sim$ 10 times larger sample of NSMs host galaxies observed at $z\leq1$ should be adequate to derive clues on the DTD parameters.

\section{Discussion} \label{sec:Discussion}
\subsection {Mass Distribution of the NSM hosts}
\begin{figure}
    \centering
    \includegraphics[trim={0cm 0cm 0cm 0cm}, clip, width=0.5\textwidth]{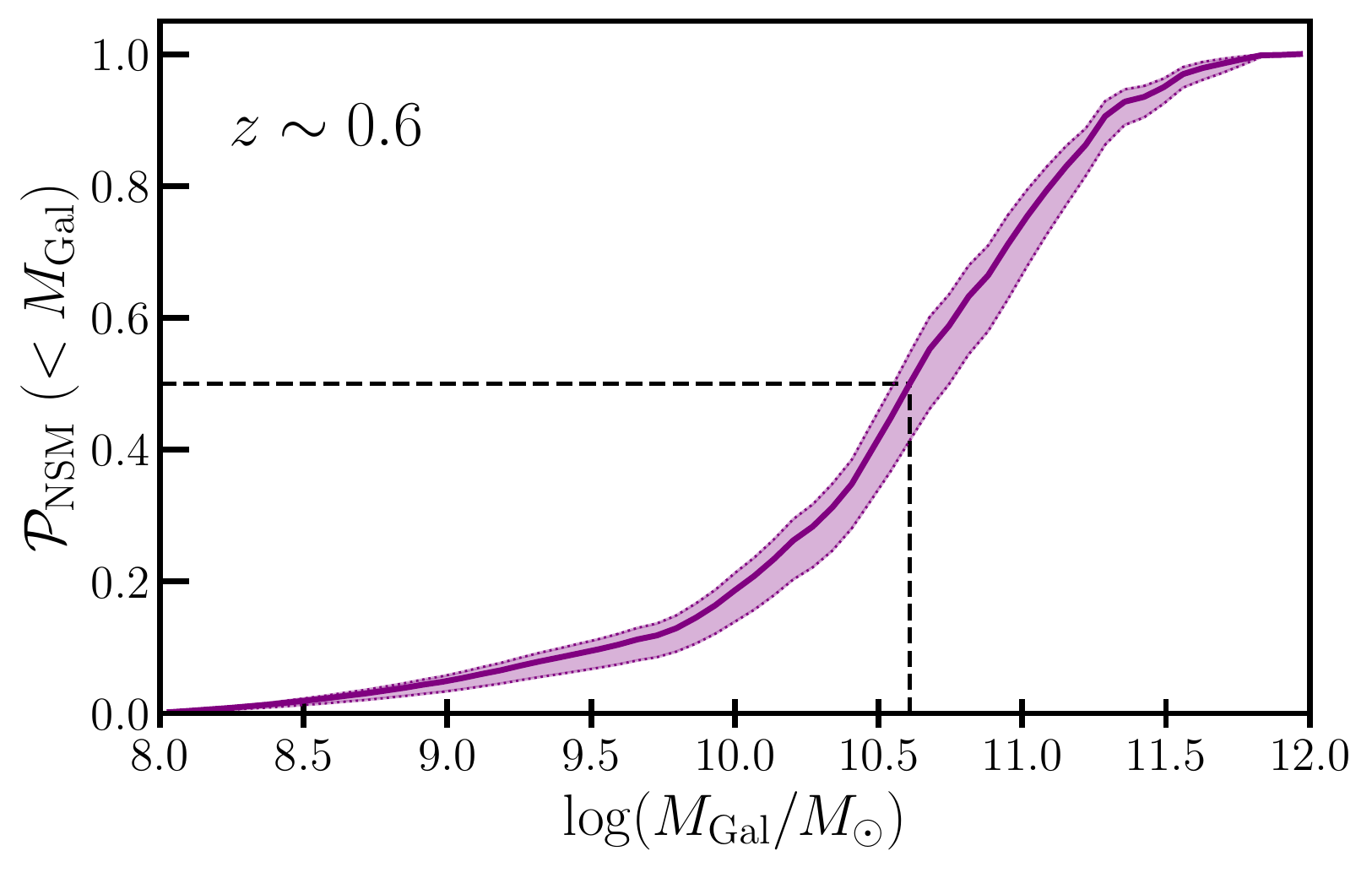}
    \caption{Mean (solid line) and dispersion (shaded area) of the cumulative distributions of $\mathcal{P}_{\rm NSM}(M_{\rm Gal})$ at $z=0.6$ for different DTDs with 30 different combinations of $A_{\rm min}$ and [$\gamma,\rho$] (same combinations of Fig. \ref{fig:Probabilityatz0}). The dash line shows the value of $M_{\rm Gal}$ at which the mean cumulative distribution reaches the $50 \%$ (i.e. $\log(M_{\rm Gal}/\sunmass)=10.6$).  }
    \label{fig:Probabilityatz0.6}
\end{figure}

As presented in Section \ref{sec:TheSampleofGalaxies}, our synthetic sample of galaxies has been obtained by merging the high-mass ($M_{\rm Gal}>10^{10}$ \sunmass\ ) galaxies of \citet{Abramson2016ApJ...832....7A} with low-mass ($10^8 < M_{\rm Gal}<10^{10}$ \sunmass\ ) ones generated by us. This low-mass extension of the \citet[]{Abramson2016ApJ...832....7A} sample has been made because even if the contribution to the cosmic SFRD from this low-mass objects is likely small, their number is much larger than that of massive galaxies and so their contribution to the NSM events may be important. Now, we estimate their contribution to the cosmic rate of NSMs.\\
We compute the fraction of NSMs hosted by galaxies with $M_{\rm Gal}<10^{10}$ \sunmass\ in our mock Universe as the ratio between the rate of NSM ($R_{\rm NSM}(z)$) in low-mass objects and the total one. This fraction depends on the DTD: when a steep DTD ($\beta=-3$) is assumed the contribution of low-mass galaxies to the rate of NSMs is higher than in the case of flat DTDs ($\beta=-1$). This trend is caused by two combined factors: i) in our picture, at $z=0$, galaxies with $M_{\rm Gal}<10^{10}$ contribute $\sim 60 \%$ to the total star formation of the local Universe; ii) steeper DTDs follow closely the SFR and so the contribution from galaxies of different masses to the rate of NSMs approaches their contribution to the star formation rate. In fact in our models with steep DTD (i.e. with $\beta=-3$) $\sim 50 \%$ of NSMs that explode at $z=0$ occur in low-mass galaxies. By contrast, in the case of a flat DTD (i.e. $\beta=-1$) only $\sim35 \%$ of NSMs are hosted by galaxies with $M_{\rm Gal}<10^{10}$ \sunmass\ .\\
The fraction of NSMs in low-mass galaxies decreases when we move to higher redshifts. At $z=0.5$, galaxies with ($M_{\rm Gal}<10^{10}$ \sunmass\
contribute $\sim20 \%$ to the total rate of NSMs irrespective of the DTD. This value is even lower at $z=1$ where only the $\sim 13\%$ of NSMs occur in low-mass galaxies. However, their contribution increases again at $z \gtrsim 2 $ where the future massive galaxies ($M_{\rm Gal}>10^{10}$ \sunmass\ at $z=0$) are building up stellar mass at the dawn of the cosmic star formation.\\
With our synthetic universe, we are also able to make some predictions on future observations of SGRBs and their host galaxies. We compute the probability ($\mathcal{P}_{\rm NSM}(M_{\rm Gal})$) that a SGRB observed at $z$ is hosted by a galaxy of mass $M_{\rm Gal}$ as:
\begin{equation}
    \mathcal{P}_{\rm NSM}^{j} = c \times \sum^{i}_{M_{\rm Gal}^i \in ]M_1,M_2]_j} R_{\rm NSM}^i(z)
\end{equation}
where $c$ is a normalisation constant such that $ \sum_j \mathcal{P}_{\rm NSM}^j=1$.\\
In Fig. \ref{fig:Probabilityatz0} we plot the distribution of $\mathcal{P}_{\rm NSM}(M_{\rm Gal})$ at redshift $z=0$ for DTDs that assume different $A_{\rm min}$ (columns), different tuples [$\gamma , \rho$] (rows), and different values of $\beta$ (colours). The distributions appear quite wide, with characteristics that vary with the fraction of prompt events in the DTD. In general, the distributions obtained with $\beta=-2$ and $-3$ are very similar for all parameters combinations, while the cases $\beta=-1$ appear systematically more populated at the high masses end. This feature is less and less apparent as the fractions of \textit{prompt} events in the DTD decreases (from the top left to bottom right panel). In other words, we expect that if the DTD is abundant in prompt events (i.e. steep $\beta$, small $A_{\rm min}$), the mass distribution of the SGRB hosts is well populated around  $10^{10}$ \sunmass.\\
The distributions shown in Fig. \ref{fig:Probabilityatz0} result from the galaxies GSMF, and from the interplay between the SFR of galaxies of different masses and the fraction of \prompt\ events in the DTD. The probability of the occurrence of a NSM event is proportional to the mass of the galaxy and to the current SFR through the $\prompt$ component. The higher this component, the more sensitive is the probability of the event to the current SFR, and the galaxy mass is relatively less important. Conversely, for DTDs with a smaller $\prompt$ fraction, the probability of the event is more sensitive to the value of the galaxy mass.\\
Fig. \ref{fig:Probabilityatz0.6} shows the cumulative mass distribution of the hosts of NSMs at $z=0.6$. At higher redshift all our DTDs produce virtually identical distributions for the expected host galaxy mass. In our mock Universe, at $z=0.6$ low-mass galaxies (i.e. $M_{\rm Gal}<10^{10}$ \sunmass) account only for the $25\%$ of the NSMs. A similar contribution comes from galaxies with $M_{\rm Gal}>10^{11}$ \sunmass. At this redshift, NSMs are more likely to be hosted ($\sim 50\%$) by galaxies between $10^{10}$ \sunmass\ and $10^{11}$ \sunmass\ irrespective of the DTD.\\
We notice that our modelling does not account for galaxy mergers by construction. To do this, we should adopt a completely different approach, e.g. couple our formalism to theoretical models of galaxy formation. Nonetheless, we remark that the distribution of the NSM rate over the galaxy population in a redshift slice depends on the mass distribution of the galaxies and on the stellar age distribution within each object in the redshift slice. As long as our formalism reproduces these properties, we catch the main features of the NSM hosts. On the other hand, including merging would shift the average mass of NSM host to the left, e.g. by $\simeq$ 0.3 dex, if one assumes that every galaxy had a single major merger in the last $\sim$6 Gyr. 
We will check the effect of galaxy mergers  quantitatively in a forthcoming paper where we  will substitute our mock Universe with that of Illustris simulations.

\subsection{Comparisons with the literature}
Previous works have presented models for the fraction of NSMs in galaxies of different type. In the following we compare our results to those in \citet{Artale2020MNRAS.491.3419A, Molero2021,Santoliquido2022MNRAS.516.3297S} and \citet{Zevin2022ApJ...940L..18Z}.\\\\
\noindent
\citet{Artale2020MNRAS.491.3419A} have analysed the redshift evolution of the rate of NSM in \passive\ and \starf\ galaxies classified as such according to their sSFR being respectively lower and higher than $10^{-10}$yr$^{-1}$. They used a population-synthesis simulation combined with galaxy catalogues from the EAGLE suite \citep{Schaye2015MNRAS.446..521S}. They found that in the local Universe (at $z\leq 0.1$) the contribution to the rate of NSM from \passive\ galaxies is greater than that one from SF galaxies, while at $z > 1$, the vast majority of mergers occurs in SF galaxies. These results are consistent with the predictions of our model for a DTD with $A_{\rm min}=1$ $R_{\odot}$ (lower left panel of Fig. \ref{fig:fractionsvsz}). 
The authors, also provide the predicted mass distribution of the hosts of NSMs at $z=0.1$, $1$, $2$, $6$ (Fig. 2 of \citet{Artale2020MNRAS.491.3419A}). We compared our distribution of $\mathcal{P}_{\rm NSM}(M_{\rm Gal})$ for $\beta=-1$ at $z=0.1$ and 1 (not shown here) to the corresponding curves in \citet{Artale2020MNRAS.491.3419A}, and found a fair agreement between the two sets of models.\\

\noindent
\citet{Molero2021} have computed the rate of NSM in elliptical, spiral, and irregular galaxies, and their contribution to the total rate in different cosmological scenarios. They test two options for the distribution of the delay times: a constant gravitational delay of 10 Myr added to the evolutionary lifetime of the secondary star and a DTD with $\beta=-0.9$. The different approach to the modelling of the evolution of galaxies in the Universe weakens the comparison between our and their results. However, we remark that in all the cosmological scenarios that they have tested, when a DTD $\propto t^{-0.9}$ is assumed their models predict that, in the local Universe, almost the $100 \%$ of NSM are hosted by SF galaxies (spirals + irregulars). However, as pointed out by \citet{Fong2022barXiv220601764N}, at $z\sim 0$ \passive\ galaxies contribute $\sim 40 \%$ of the rate of NSMs. Our models with $\beta =-1$ (DTD similar to the one used by \citet{Molero2021} $\propto t^{-0.9}$ ) predict that $\sim 30-40 \%$ of NSMs in the local Universe are hosted by \passive\ galaxies, in agreement with the results of \citet[]{Fong2022barXiv220601764N}.
Therefore the different result on the properties of the NSM hosts reflects the dissimilar approach used to model the evolution of the galaxy population in the Universe.\\
Another difference concerns the value of $\alpha_{\rm NSM}$ which \citet{Molero2021} found of  $\sim 0.05 -0.06$, i.e. more than 10 times larger than found here. Part of the discrepancy is due to their use of the old value of the cosmic NSM rate at zero redshift \citep{Abbott2020}, that was a factor of $\sim$ 3 larger than what adopted here. \\\\
\noindent
\citet{Santoliquido2022MNRAS.516.3297S} studied the redshift evolution of the merger rate density of compact objects and the properties of the binary systems hosts. They use a BPS code to model the delay times of compact objects, while the evolving galaxy population in the Universe obeys empirical fundamental scaling relations. These prescriptions determine the formation galaxy of the binary compact objects, while the host galaxy of the final mergers is found by applying a merger tree to the results of the EAGLE cosmological simulation \citep{Schaye2015MNRAS.446..521S}. The authors present the results obtained with different assumptions, in particular concerning the efficiency of the Common Envelope $\alpha_{\rm CE}$. In spite of the different approach, it is worth comparing our results to theirs for the merging of binary neutron stars.\\
The cosmic evolution of the NSM rate in \citet{Santoliquido2022MNRAS.516.3297S} models is broadly consistent with ours: the rate increases from redshift z=0 up to $z \simeq 2$ of a factor which varies between $\sim$ 1.5 and $\sim$ 8 depending on the assumptions for the galaxy population and the $\alpha_{\rm CE}$ parameter. Our models present a similar variation for the explored range of parameters characterising the DTD, i.e. $\beta$ and $A_{\rm min}$. \\
\citet[]{Santoliquido2022MNRAS.516.3297S} also found that in the local Universe the fraction of NSM hosted by \passive\ galaxies strongly depends on the classification criteria. Specifically, they find that $\sim 50-60 \%$ of local NSMs occur in \passive\ galaxies if these are defined as objects with sSFR $<10^{-10}$ yr$^{-1}$ (same criteria used by \citet{Artale2020MNRAS.491.3419A}). On the other hand, if \passive\ galaxies are selected as those with a SFR 1 dex below the star-forming main sequence (similar to the criterion of sSFR$<10^{-11}$ yr$^{-1}$, right column of Fig. \ref{fig:fractionsvsz}), only $\sim 5-10 \%$ of NSMs are hosted by \passive\ galaxies. Their results are similar to those obtained with our modelling for a DTD with $\beta=-2,-3$ and $A_{\rm min}<1$ $R_{\odot}$ (see bottom row of Fig. \ref{fig:fracvsz}).\\\\
\noindent
\citet{Zevin2022ApJ...940L..18Z} have placed constraints on the DTD of SGRBs using observed SGRBs with host association. Assuming a pure power-law DTD, they used a sample 69 SGRBs with hosts from \citet{Fong2022aarXiv220601763F} to infer the index of the distribution. They conclude that SGRBs/NSMs have a DTD with a power-law index $\sim -1.8$, with a $90\%$ credible interval that ranges form $-1.4$ and $-2.1$.  The conclusions of \citet{Zevin2022ApJ...940L..18Z} are agreement with our results. In particular, we found that the fraction of SGRBs observed in \starf\ galaxies advocates for DTD models with $\beta=-2/-3$, which at long delays are well fit by a power-law with a slope of $-1.5/-2.0$).

\begin{figure}
    \centering
    \includegraphics[trim={0cm 0cm 0cm 0cm}, clip, width=0.5\textwidth]{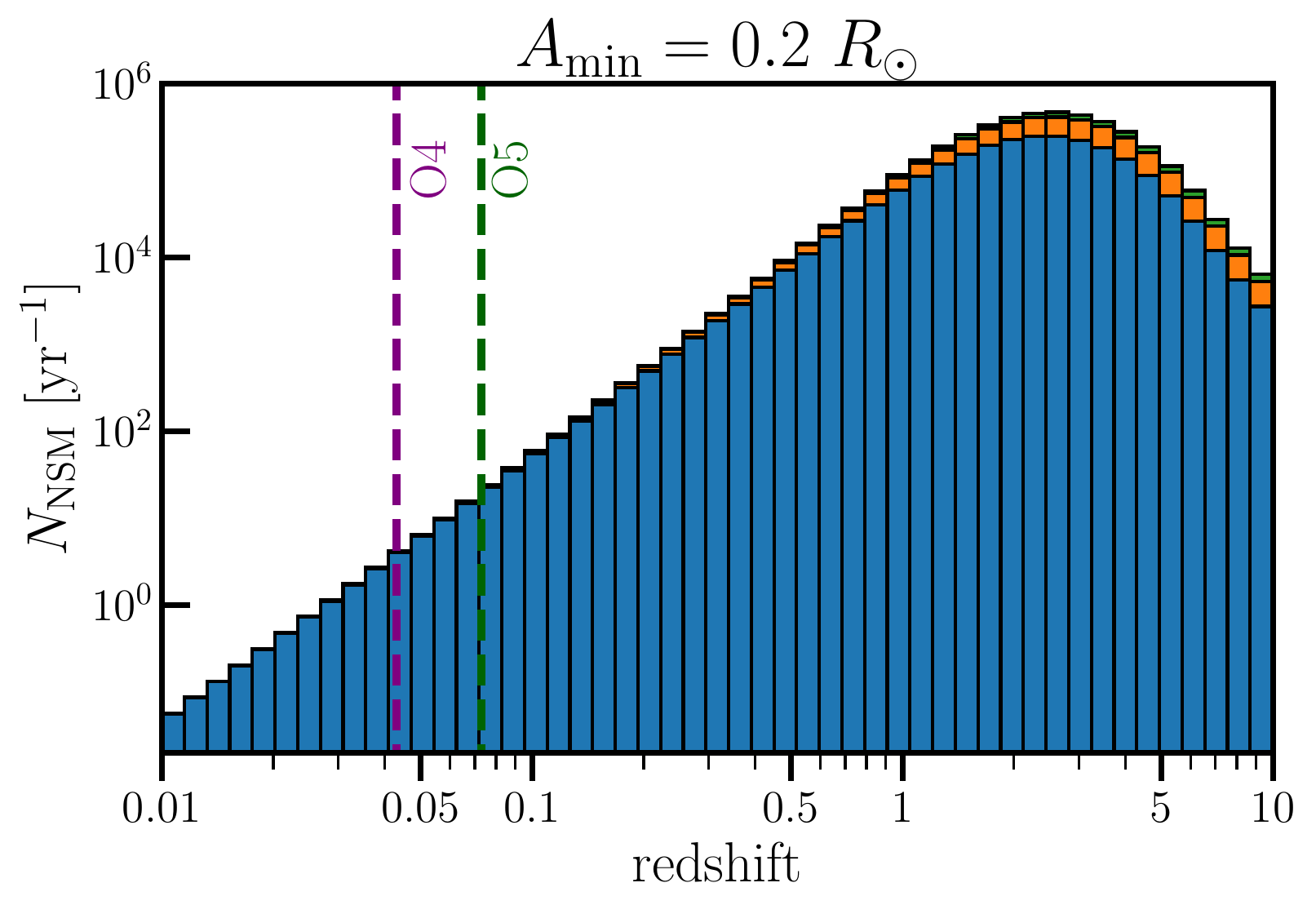}
    \caption{Redshift distribution of the number of NSMs per year in our mock Universe for a DTD with $A_{\rm min}=0.2$ $R_{\odot}$ and $\beta=-1$ (blue), $\beta=-2$ (orange), $\beta=-3$ (green). With vertical dotted lines are reported the redshift limits for O4 (purple) and O5 (dark green) observing runs \citep[see][]{LigoVirgoColl2020LRR....23....3A}. }
    \label{fig:O4O5}
\end{figure}

\section{Conclusions} \label{sec:conclusions}
In this paper, we presented a detailed investigation of possible observational facts that can allow us to constrain the delay time distribution of neutron star mergers. To do that we first developed a mock Universe composed of a sample of galaxies that fulfils three major observational constraints: i) the star formation rate density in the Universe by \citet{MD2014}; ii) the galaxy stellar mass function of nearby galaxies derived by \citet{Peng2010ApJ...721..193P}; iii) the star-forming main sequence of galaxies \citep{Renzini2015ApJ...801L..29R}. The galaxies in our mock universe 
undergo a SFH described by a lognormal function with two parameters, as proposed by \citetalias{G13}, and include massive ($M_{\rm Gal}>10^{10}$ \sunmass) galaxies \citep{Abramson2016ApJ...832....7A} to which we add low-mass galaxies down to $M_{\rm Gal} = 10^8$ $\sunmass$. 
The redshift evolution of the rate of neutron star mergers is then computed by convolving the SFH of each galaxy with the parametric delay time distributions derived by \citet[]{Greggio2021}. The shape 
of these DTDs depends on four parameters: $\beta$, $A_{\rm min}$, $\gamma$ and $\rho$, representing respectively the slope of the distribution of the separations and its minimum value, the slope of the distribution of the total masses and that of the distribution of the eccentricities, all parameters characteristic of the binary neutron star systems at birth.\\
The local rate of binary neutron star mergers of \citet{2021ApJ...913L...7A} is used to calibrate the global model for the cosmic NSM rate, yielding a value for the realization probability of the NSM evolutionary channel. In addition, our modelling allows us to present the expected relations between the rate of NSM and the properties of the host galaxy.\\
Our main results are as follows:
\begin{list}{--}{} 
    \item Between $z=0$ and $z=2$ the rate of NSMs increases with a slope that depends on the parameters of the DTD (mainly on $\beta$ and $A_{\rm min}$). The difference between the highest rate (i.e. at $z \simeq 2$) and the local one increases with the fraction of prompt events characterizing the DTD. Assuming that NSMs are the progenitors of SGRBs, we compare our theoretical curves with the redshift evolution of the rate of SGRBs computed by \citet{Ghirlanda2016}. This constraint favours DTD with $\beta=-1$, as found in \citet{Greggio2021}.\\
    \item Adopting the current estimate for the rate of NSMs in the local Universe, we estimate that the fraction of neutron star progenitors living in binary systems with the right characteristics to lead to a NSM within a Hubble time ranges  
    between $\sim0.1 \%$ and $\sim1 \%$. This large uncertainty is a direct consequence of the uncertainty affecting the local rate of NSM mergers computed by \citet[]{2021ApJ...913L...7A}. We find that the most probable value for the realization probability is of $0.3 \%$ with a weak dependence on the DTD parameters $\beta$ and $A_{\rm min}$ (see Fig. \ref{fig:alphas}).\\\\
    
    \item We find a dramatic variation of the fraction of NSMs hosted by $\starf$ galaxies among models with different fractions of prompt events, mainly controlled by the $\beta$ and $A_{\rm min}$ parameters. In addition, at a given DTD model, the fraction of NSM hosted by $\starf$ galaxies is found very sensitive to the criterion used to divide galaxies in $\starf$ and $\passive$. \citet{Fong2022barXiv220601764N} found that $\sim 85\%$ of SGRBs are observed in \starf\ galaxies. Adopting the same classification criterion, we find that the fraction of SGRBs observed in \starf\ galaxies favours DTDs with at least $\sim40\%$ of mergers within $100$ Myr ($\beta=-2/-3$). These figures refer to $z \simeq 0.6$ which is the average redshift of the host galaxies in the \citet{Fong2022barXiv220601764N} sample.\\\\
    \item The evolution with redshift of the fraction of NSMs 
    in \starf\ galaxies depends on the criterion used to define them and on the DTD. Current empirical estimates of this trend \citep{Fong2022barXiv220601764N} are affected by large uncertainties caused by the poor statistics so that all our models are compatible with the data. Larger datasets ($\sim 600$ events with $z \leq 1$) will allow exploiting this trend to constrain the DTD (see Appendix \ref{app:SGRBstatistic}).\\\\
    \item The combined effect of the DTD and GSMF is such that in the local Universe we expect a different distribution of host mass for different $\beta$’s. In particular, steeper DTDs ($\beta=-2/-3$) generate more events in lower mass galaxies, compared with flatter ones ($\beta=-1$). This difference rapidly vanishes as $z$ increases, and at $z\sim0.6$ the median host mass around $\log(M_{\rm Gal}/\sunmass)= 10.6$ virtually irrespective of the DTD shape.\\\\
    \end{list}
    In Fig. \ref{fig:O4O5} we show the redshift distribution of the number of NSMs per year in our mock Universe. The steep increase with redshift of the sampled volume dominates over the increase of the volume density of the rate of NSM, so the redshift distribution of the rate of NSM is almost independent of the shape of the DTD. Only at $z \gtrsim 1$ does the rate of events show a higher value for those DTDs with a larger fraction of short delay times, with a difference of $60 \%$ at z=1 increasing to more than $100 \%$ at the peak of the distribution going from $\beta=-1$ to $\beta=-2,-3$.\\
    In Fig. \ref{fig:O4O5} we also show the expected depth for the sensitivity to NSMs of the observing runs O4\footnote{Advanced LIGO + Advanced Virgo Plus + KAGRA} and O5\footnote{Advanced LIGO + Advanced Virgo Plus + KAGRA + LIGO-India (from 2025)}. In the O4 run \citep[see][]{LigoVirgoColl2020LRR....23....3A}, Advanced LIGO will be the deepest instrument of the network sensitive to NSMs occurring at a distance smaller than 160-190 Mpc, which corresponds to a $z_{\rm O4}\simeq0.043$. In one year, in our mock Universe $\sim8^{+11}_{-6}$ NSMs explode within the redshift limit of O4 ($z_{\rm O4}$). The preliminary plans for O5 indicate a target for the sensitivity to NSMs events up to a distance of $330$ Mpc that corresponds to $z_{\rm O5}\simeq 0.073$. In this case, we expect a number of $\sim44^{+65}_{-33}$ in one year.\\\\
\noindent
SGRBs with host association seems to be the most reliable short-term observational proxy to put some constraints on the DTD of NSMs. On the other hand, the third-generation of gravitational-wave detectors network may provide samples of gravitational waves events large enough to yield a complementary way to determine the DTD of NSMs \citep[see][]{Safarzadeh_I_2019ApJ...878L..12S,Safarzadeh_II_2019ApJ...878L..13S,Safarzadeh_III_2019ApJ...878L..14S, McCarthy2020MNRAS.499.5220M}.\\
We remark that the figures presented here depend on our description of the galaxy population in the Universe. Therefore our predictions will be to some extent subject to revision with different constraints, especially regarding the SFH of galaxies. Nevertheless, the most promising data to constrain the DTD of NSM appears to be the fraction of events occurring in \starf\ galaxies, and its evolution with redshift. 
\section*{Data availability}
The data underlying this article will be shared on reasonable request to the corresponding author.

\section*{Acknowledgements}
We thank an anonymous referee for his/her comments which improved the quality of this paper. We are indebted to Dr L. Abramson for sending us the data of the G13 galaxy sample, which are at the basis of our computations.




\bibliographystyle{mnras}
\bibliography{example} 




\appendix

\section{On the galaxy stellar mass function}\label{app:Overdensity}
\begin{figure}
    \centering
    \includegraphics[trim={0cm 0cm 0cm 0cm}, clip, width=0.5\textwidth]{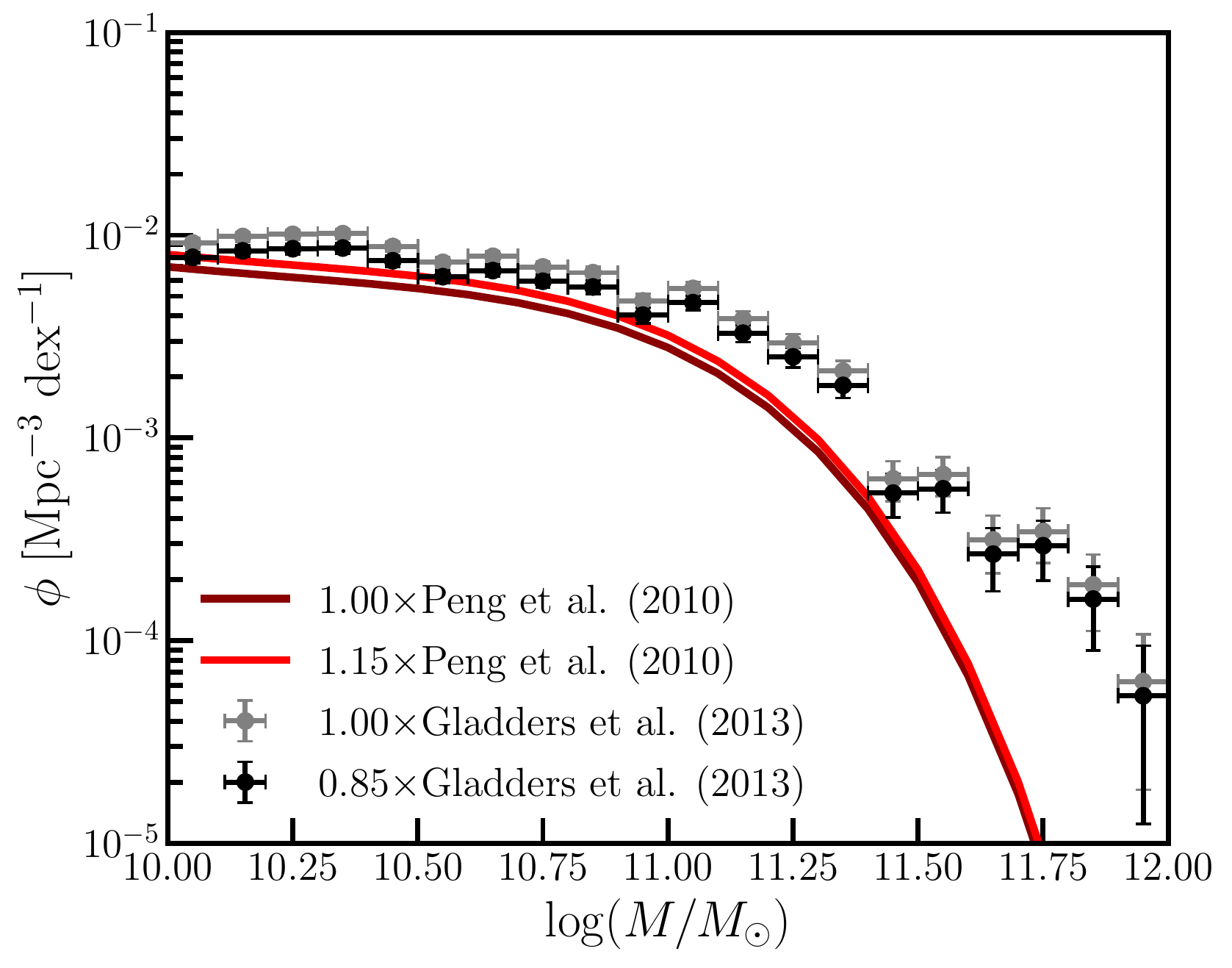}
    \caption{Original (dark red line and grey dots) and modified (red line and black dots) mass functions of \citetalias{G13} and \citetalias{Peng2010ApJ...721..193P}. The final GSMF adopted in shown in Fig. \ref{fig:MDF}. Note that we refer to the \citet{Peng2010ApJ...721..193P} GSMF rescaled to the Salpeter IMF.}
    \label{fig:MDF_alt}
\end{figure}

\begin{figure}

    \centering
    \includegraphics[trim={0cm 0cm 0cm 0cm}, clip, width=0.5\textwidth]{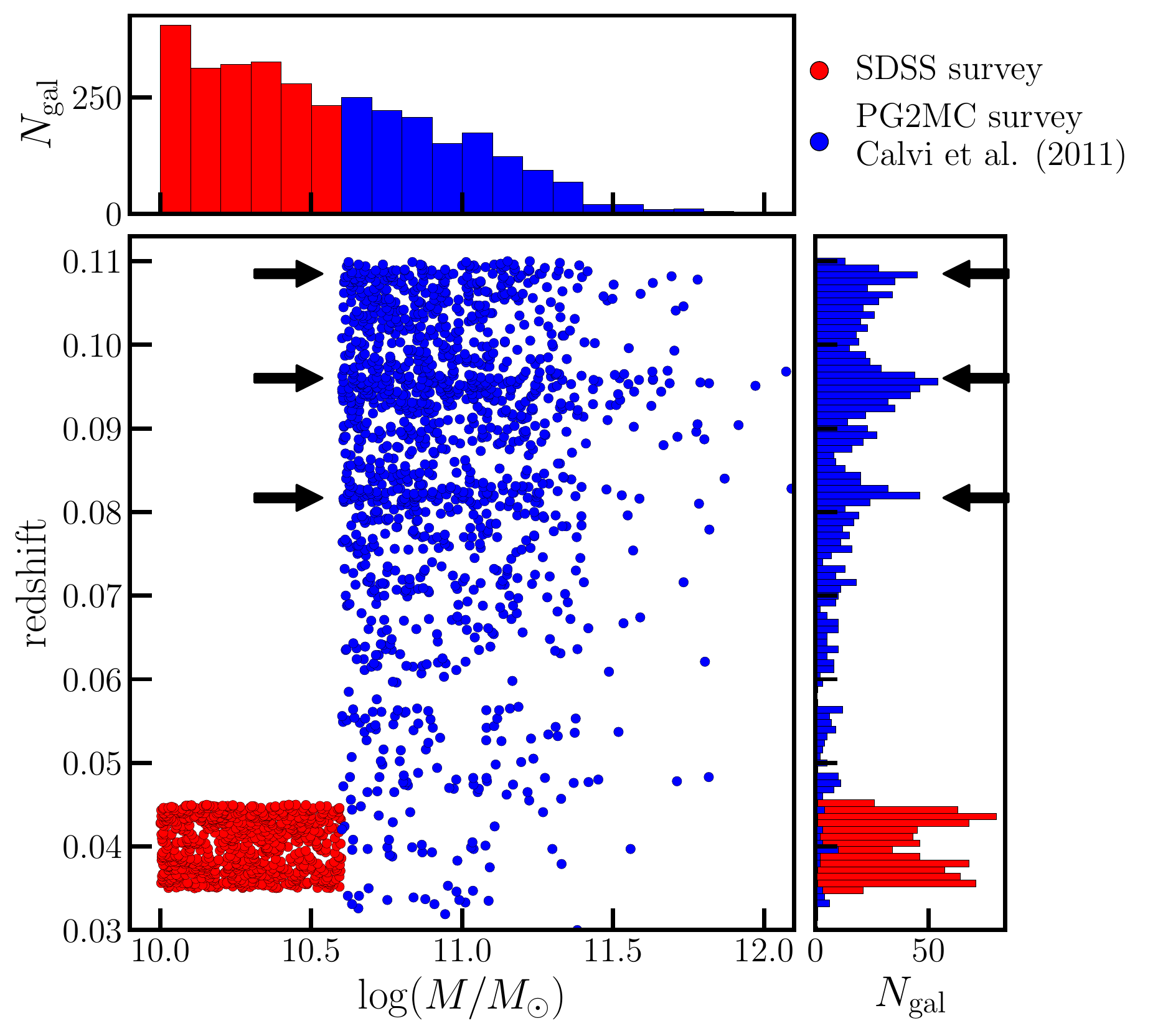}
    \caption{Redshift distributions of the two sub-samples that compose the one of \citetalias{G13}. The SDSS (red dots) survey has a smaller volume compared to the PG2MC (blue dots) one. Black arrows indicate the three over-densities of galaxies present in the PG2MC sample.}
    \label{fig:RedDistr}
\end{figure}

As discussed in Section \ref{sec:TheSampleofGalaxies}, the sample of galaxies of \citetalias{G13} shows an over-density of objects at $M_{\rm Gal}>10^{11}$ $\sunmass$. In Fig. \ref{fig:MDF_alt}
we show the original sample of \citetalias{G13} (plotted in grey) and the GSMF of \citet{Peng2010ApJ...721..193P} (dark red solid line); one can see that an over-density is present in all the mass bins. As reported in \citetalias{G13}, their sample of galaxies is composed of two sub-samples. The first has been taken from the Padova-Millennium Galaxy and Group Catalogue (PG2MC) survey \citep[see ][]{Calvi2011MNRAS.416..727C}, which covers a large volume ($0.03\leq z\leq 0.11$) and has a mass lower limit of $\num{4e10}$ $\sunmass$. The second, is from the Sloan Digital Sky Survey (SDSS) observations of the northern galactic cap, covers a smaller volume ($0.035\leq z \leq 0.045$), and has a mass limit of $\num{1e10}$ $\sunmass$. This second sample has been used to extend the PG2MC sample to the lower mass regime and has been cut at a mass of $\num{4e10}$ $\sunmass$. When constructing the GSMF we account for the different Volumes sampled by the two sub-samples. Fig. \ref{fig:RedDistr} shows the redshift distribution of the two sub-samples. It appears that the PG2MC sample shows three clear over-densities of points at specific redshifts: $z\sim 0.08$, $z\sim 0.09$, and $z\sim 0.11$, indicated with black arrows. These features may indicate that a fraction of the galaxies contained in the survey belong to galaxy clusters. This could be responsible for the over-density of the \citetalias{G13} GSMF with respect to the one in \citetalias{Peng2010ApJ...721..193P}.\\
We estimated that, on average, the number of galaxies per mass bin of \citetalias{G13} differs from the one of \citetalias{Peng2010ApJ...721..193P} as follows:
\begin{equation}
    \mathcal{N}_{\rm P10} = 0.65\times \mathcal{N}_{\rm G13} \ \ .
\end{equation}
However, with such a great reduction of galaxies, the SFRD of the \citetalias{G13} sample is not in agreement with the one determined by \citet{MD2014}. In addition, we notice that part of the discrepancy between the P10 and G13 GSMFs may come from the different ways in which the return fraction is included in the evaluation of the current galaxy mass. We thus adopt the following compromise to combine the GSMFs: increasing the GSMF of \citetalias{Peng2010ApJ...721..193P} by a factor of $1.15$ and reducing the one of \citet{G13} by a factor of $0.85$. As shown in Fig. \ref{fig:MDF_alt}, this solution relieves the tension In the range $M_{\rm Gal} \lesssim 10^{11}$  $\sunmass$ and, at the same time, allows us to reproduce the SFRD of \citet{MD2014} (see Fig. \ref{fig:SFRD}).

\section{Short Gamma-Ray Bursts statistic}\label{app:SGRBstatistic}
\begin{figure}
    \centering
    \includegraphics[width=0.5\textwidth]{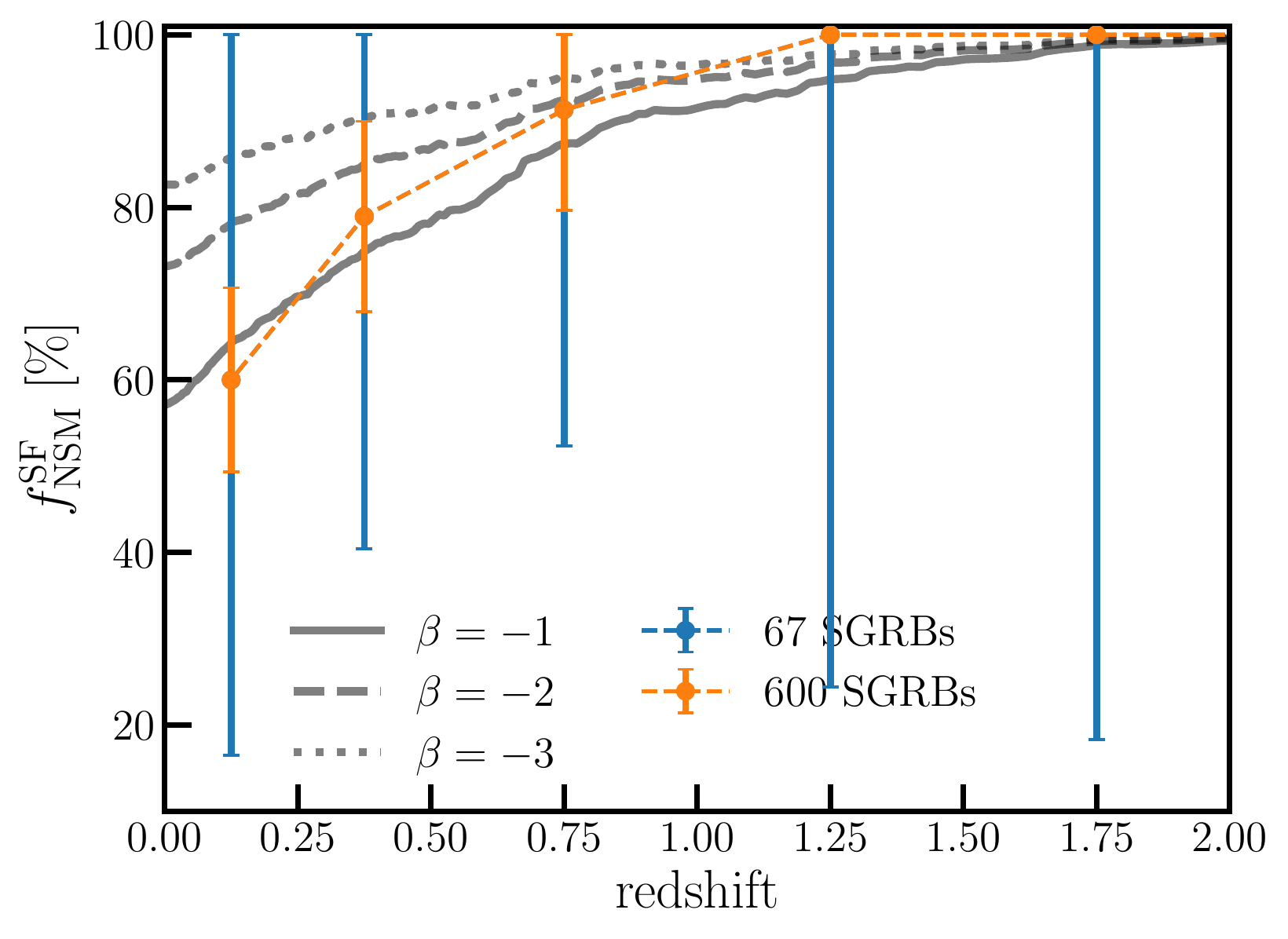}
    \caption{Redshift evolution of the fraction of NSMs hosted by SF galaxies for DTDs with $\beta=-1$ (solid line), $\beta=-2$ (dashed line), and $\beta=-3$ with $A_{\rm min}= 0.5$ $ {\rm R}_{\odot}$.  We also plot the redshift evolution of the fraction of SGRBs hosted by SF galaxies and its uncertainty from the data of \citet{Fong2022barXiv220601764N} (in blue) and from an hypothetical sample of 600 SGRBs with $z \leq 1$ ($\sim 10$ times larger than the current one) with host association (in orange).}
    \label{fig:B1}
\end{figure}

As discussed in Section \ref{sec:fractions}, the observational values of the redshift evolution of SGRBs hosted by SF galaxies are affected by large uncertainties. These are caused by the shot noise associated with the number of SGRBs observed in a certain redshift bin.\\
To compute the observational values and their uncertainties we retrieved for each SGRB contained in the sample of \citet{Fong2022barXiv220601764N} its redshift ($z$) and the type of the host (\passive\ or SF). In Table \ref{tab:SGRB} we report, for six redshift bins: the total number of SGRBs ($N_{\rm TOT}$), the number of SGRBs hosted by SF galaxies ($N_{\rm SF}$), and the fraction $f_{\rm SF} = N_{\rm SF}/N_{\rm TOT}$ with its uncertainties. 
We evaluate the uncertainties with the relation:
\begin{equation}
    \Delta_{f_{\rm SF}} = f_{\rm SF} \times \left( \frac{\Delta_{\rm TOT}}{N_{\rm TOT}} + \frac{\Delta_{\rm SF}}{N_{\rm SF}}  \right)
\end{equation}
where $\Delta_{\rm TOT}$ and $\Delta_{\rm SF}$ are the Poissonian errors associated to
$N_{\rm TOT}$ and $N_{\rm SF}$ respectively, i.e. \\
\begin{gather*}
\Delta_{\rm TOT} =  \sqrt{N_{\rm TOT}} \hspace{1cm} \Delta_{\rm SF} =  \sqrt{N_{\rm SF}}
\end{gather*}
We also computed the uncertainties of an hypothetical sample of 600 SGRBs with host association at $z\leq 1$ ($\sim 10$ times larger than the current one). To compute this, we kept fixed redshift evolution of $f_{\rm NSM}^{\rm SF}$ and we simply increase the number SGRBs associated with SF or \passive\ galaxies in each redshift bin, up to $z=1$. On Fig. \ref{fig:B1} we show the uncertainties of the current sample (in blue), and those of a $\sim 10$ times larger sample (in orange). By improving the statistics of SGRBs with host association at low redshift ($z\leq1$) by one order of magnitude we could be able to discriminate between DTD with different $\beta$s. This is unfeasible with the current uncertainties.

\begin{table}
	\centering
	\caption{Total number of SGRBs ($N_{\rm TOT}$), number of them hosted by SF galaxies ($N_{\rm SF}$), and the fraction $f_{\rm SF}$ with its uncertainties in six redshift bins from the sample of \citet{Fong2022barXiv220601764N}.}
	\label{tab:SGRB}
	\renewcommand{\arraystretch}{1.5}
	\begin{tabular}{lccc} 
		\hline
		  & $N_{\rm TOT}$ & $N_{\rm SF}$ & $f_{\rm SF}$\\
		\hline
		$2.5 \leq z < 2.0$  & $2 $ & $2 $ & $1.00_{-1.00}^{+0.00}$\\
		$2.0 \leq z < 1.5$  & $6 $ & $6 $ & $1.00_{-0.82}^{+0.00}$\\
		$1.5 \leq z < 1.0$  & $7 $ & $7 $ & $1.00_{-0.76}^{+0.00}$\\
		$1.0 \leq z < 0.5$  & $23$ & $21$ & $0.91_{-0.39}^{+0.09}$\\
	    $0.5 \leq z < 0.25$ & $19$ & $15$ & $0.79_{-0.38}^{+0.21}$\\
	    $z \leq 0.25$       & $10$ & $6 $ & $0.60_{-0.43}^{+0.40}$\\
		\hline
	\end{tabular}
\end{table}

\bsp	
\label{lastpage}
\end{document}